\documentclass[apjl]{emulateapj} 
\usepackage[pass,letterpaper]{geometry}
\usepackage{hyperref}

\journalinfo{Submitted to the Astrophysical Journal}
\submitted{Received 2015 April XX; accepted 2015 YYYYYYY XX; published 2015 YYYYYYY XX}

\shorttitle{Kinematics and Chemistry of Reticulum~2 and Horologium~1}
\shortauthors{Koposov et al}

\begin{document}

\title{Kinematics and Chemistry of recently discovered Reticulum~2 and Horologium~1 dwarf galaxies}

\author{Sergey E. Koposov}
\affiliation{Institute of Astronomy, Madingley Road, Cambridge CB3 0HA, UK}
\author{Andrew R.\ Casey}
\affiliation{Institute of Astronomy, Madingley Road, Cambridge CB3 0HA, UK}
\author{Vasily Belokurov}
\affiliation{Institute of Astronomy, Madingley Road, Cambridge CB3 0HA, UK}
\author{James R. Lewis}
\affiliation{Institute of Astronomy, Madingley Road, Cambridge CB3 0HA, UK}
\author{Gerard Gilmore}
\affiliation{Institute of Astronomy, Madingley Road, Cambridge CB3 0HA, UK}
\author{Clare Worley}
\affiliation{Institute of Astronomy, Madingley Road, Cambridge CB3 0HA, UK}
\author{Anna Hourihane}
\affiliation{Institute of Astronomy, Madingley Road, Cambridge CB3 0HA, UK}


\author{T. Bensby}
\affiliation{Lund Observatory, Department of Astronomy and Theoretical Physics, Box 43, SE-221 00 Lund, Sweden}

\author{A. Bragaglia}
\affiliation{INAF - Osservatorio Astronomico di Bologna, via Ranzani 1, 40127, Bologna,
Italy}

\author{M. Bergemann}
\affiliation{Max-Planck Institut f\"{u}r Astronomie, K\"{o}nigstuhl 17, 69117 Heidelberg, Germany}

\author{G. Carraro}
\affiliation{European Southern Observatory, Alonso de Cordova 3107 Vitacura, Santiago de Chile, Chile}

\author{E. Flaccomio}
\affiliation{INAF - Osservatorio Astronomico di Palermo, Piazza del Parlamento 1, 90134, Palermo, Italy}

\author{U. Heiter}
\affiliation{Department of Physics and Astronomy, Uppsala University, Box 516, SE-751 20 Uppsala, Sweden}

\author{V. Hill}
\affiliation{Laboratoire Lagrange (UMR7293), Universit\'e de Nice Sophia Antipolis, CNRS,Observatoire de la C\^ote d'Azur, CS 34229,F-06304 Nice cedex 4, France}

\author{P. Jofre}
\affiliation{Institute of Astronomy, Madingley Road, Cambridge CB3 0HA, UK}

\author{P. de Laverny}
\affiliation{Laboratoire Lagrange (UMR7293), Universit\'e de Nice Sophia Antipolis, CNRS,Observatoire de la C\^ote d'Azur, CS 34229,F-06304 Nice cedex 4, France}

\author{L. Monaco}
\affiliation{Departamento de Ciencias F\'{i}sicas, Universidad Andr\'es Bello, Rep\'ublica 220, 837-0134 Santiago, Chile}

\author{L. Sbordone}
\affiliation{Millennium Institute of Astrophysics, Av. Vicu\~{n}a Mackenna 4860, 782-0436 Macul, Santiago, Chile}
\affiliation{Pontificia Universidad Cat\'{o}lica de Chile, Av. Vicu\~{n}a Mackenna 4860, 782-0436 Macul, Santiago, Chile}

\author{\v{S}. Mikolaitis}
\affiliation{Institute of Theoretical Physics and Astronomy, Vilnius University, A. Go\v{s}tauto 12, LT-01108 Vilnius}

\author{N. Ryde}
\affiliation{Lund Observatory, Department of Astronomy and Theoretical Physics, Box 43, SE-221 00 Lund, Sweden}


\email{koposov@ast.cam.ac.uk,arc@ast.cam.ac.uk,vasily@ast.cam.ac.uk}

\begin{abstract}
Photometry alone is not sufficient to unambiguously distinguish between ultra-faint star clusters and dwarf galaxies because
of their overlap in morphological properties. Accurate measurements of the intrinsic velocity and metallicity dispersions are required to robustly classify such objects.
Here we report on VLT/GIRAFFE spectra of candidate member stars in two recently discovered ultra-faint satellites Reticulum~2 and Horologium~1, obtained as part of the ongoing Gaia-ESO Survey. We identify 18 members in Reticulum~2 and 5 in Horologium~1. We find Reticulum~2 to have a velocity dispersion of ${3.22^{+1.64}_{-0.49}\,{\rm km\,s}^{-1}}$, implying a $M/L$ ratio of $\sim$500. We have inferred stellar parameters ($T_{\rm eff}$, $\log{g}$, [Fe/H], $[\alpha/{\rm Fe}]$) for all candidates and we find Reticulum~2 to have a mean metallicity of ${\rm [Fe/H]} = -2.46^{+0.09}_{-0.10}$, with an intrinsic dispersion of $0.29^{+0.13}_{-0.05}$\,dex, and is $\alpha$-enhanced to the level of $[\alpha/{\rm Fe}] \sim  0.4$\,dex. We conclude that Reticulum~2 is a dwarf galaxy. We also report on the serendipitous discovery of four stars in a previously unknown stellar substructure near Reticulum~2 with ${\rm [Fe/H]} \sim -2$ and ${V_{hel}\sim 220\,{\rm km\,s}^{-1}}$, far from the systemic velocity of Reticulum~2 at ${V_{hel} = 64.7^{+1.3}_{-0.8}}$\,km s$^{-1}$. For Horologium~1 we infer a velocity dispersion of $\sigma\left(V\right) =  4.9^{+2.8}_{-0.9}\,{\rm km\,s}^{-1}$ and a consequent $M/L$ ratio of $\sim$600, leading us to conclude that Horologium~1 is also a dwarf galaxy. Horologium~1 is slightly more metal-poor than Reticulum~2 (${\rm [Fe/H]} = -2.76^{+0.1}_{-0.1}$) and is similarly $\alpha$-enhanced: ${[\alpha/{\rm Fe}] \sim 0.3 }$\,dex. Despite a large error-bar, we also measure a significant spread of metallicities   ($\sigma\left({\rm [Fe/H]}\right) = 0.17^{+0.20}_{-0.03}$\,dex) which strengthen the evidence that Horologium~1 is indeed a dwarf galaxy. The line-of-sight
velocity of Reticulum~2 is offset by some 100~km\,s$^{-1}$ from the
prediction of the orbital velocity of the LMC, thus making its
association with the Cloud uncertain. 
However, at the location of Horologium~1,
both the backward integrated LMC's orbit and the LMC's halo are
predicted to have radial velocities similar to that of the
dwarf.
Therefore, it is very likely that Horologium~1 is or once was a member
of the Magellanic Family.
\end{abstract}

\section{Introduction}
\label{sec:introduction}

The total inventory of satellites associated with the Milky Way remains incomplete. This is particularly true for the faintest systems, as observations are inherently biased towards finding and characterising intrinsically bright satellites. As a consequence, any inferences on the total satellite luminosity or mass distributions strongly depend on systematic and selection effects in the least luminous systems. Deep, uniform photometry is required to find these relics, and spectroscopy is required for proper characterisation.

Wide-field photometric surveys can be extremely successful at finding
Galactic satellites \citep{willman_2005a, willman_2005b, zucker_2006a,
  zucker_2006b, belokurov_2006, belokurov_2007a, belokurov_2008,
  belokurov_2009, belokurov_2010,
  irwin_2007,koposov_2007,walsh_2007,grillmair_2009,balbinot_2013}. The
Sloan Digital Sky Survey \citep[SDSS,][]{sdss_dr7,sdss_final} data
unveiled more than a dozen systems, opening entire new sub-fields of
astrophysics devoted to understanding these satellites and their
trailing debris \citep[see reviews by Willman 2010 and Belokurov 2013;
  see also][and references therein]{casey_2012, casey_2013,
  casey_2014,martin_2013, koposov_2013, deason_2014, deboer_2014,
  grillmair_2014, lee_2015}. Searches using the early Pan-STARRS and
VST ATLAS survey data were less successful, revealing only two new
satellites thus far \citep{belokurov_2014, laevens_2015}. More
recently however, the publicly-accessible Dark Energy Survey
\citep[DES hereafter;][]{des_2005} data has been used by two independent groups
to find at least another 9 satellites
\citep{koposov_2015,bechtol_2015}. The wide-field imaging
capabilities of the DECam have also been exploited by a number of
smaller-scale surveys of the Milky Way halo, increasing the tally of
Galactic satellites at a breakneck pace \citep{kim_2015a, kim_2015b,
  kim_2015c,martin_2015}. Many of these latest discoveries are
remarkably feeble, fainter than most known systems, rightfully earning
the name of `ultra-faint' satellites.

As more ultra-faint satellites (UFS) have being discovered, complications have arisen in trying to accurately classify them. A tenuous overlap between the effective radii and absolute magnitudes of faint globular clusters and ultra-faint dwarf galaxies has emerged. It is now crystal clear, the morphology of ultra-faint systems near the \textit{valley of ambiguity} cannot be classified from photometry alone \citep{gilmore_2007}. Kinematics and chemistry are required to distinguish between these classifications. A large dispersion in overall metallicity is representative of extended star formation in a dwarf galaxy-like environment that is massive enough to retain supernova ejecta, providing a key diagnostic for distinguishing globular clusters and dwarf galaxies \citep[e.g.,][]{willman_2012}. Indeed, spectroscopy is essential for a large number of confirmed members in order to precisely measure velocity and chemical dispersions, estimate the dark matter content, and explore the star formation histories of these ancient systems \citep[e.g.,][]{kirby_2011,tollerud_2012}.

The UFSs have been the focus of attention of Galactic archaeologists
world wide for less than a decade. Worryingly, during this short
history, many of their spectroscopically determined properties have
continued to evolve. For example, the early studies of the Bo\"{o}tes~1 dwarf spheroidal reported a velocity dispersion as high as 6.5~km
s$^{-1}$ \citep[][]{munoz_2006a,martin_2007}. However, an independent
and novel study by \citet{koposov11} revealed that the dwarf's
internal kinematics is potentially dominated by a stellar population with a
velocity dispersion as low as 2.4 km s$^{-1}$. 
Similarly, for the Segue~2 satellite, \citet{belokurov_2009} gave an estimate of 3.4~km\,s$^{-1}$ and a warning of a potential contamination from the surrounding Tri-And stream. In fact, \citet{kirby_2013} later showed the velocity of Segue~2 is consistent with zero, thus ruling out the presence of any significant amount of dark matter in the system.
Undoubtedly, robust
uncertainties on individual stellar velocity measurements are
paramount to accurate characterization of the kinematics of these
systems. Furthermore, inferences are susceptible to low-number statistics, as well as contamination from foreground stars or binary systems.

The Gaia-ESO Survey \citep{gilmore_2012,randich_2013} has been obtaining high-resolution spectra using the Very Large Telescope (VLT) in Chile since January 2012. The primary scientific goal of the Gaia-ESO Survey is to characterise the major constituents of the Galaxy, and to understand these components in the context of the Milky Way's formation history. To that end, more than 10$^5$ Milky Way stars are homogeneously targeted as part of the Gaia-ESO Survey, including all major structural components: open and globular clusters, the disk, bulge, and the halo. Understanding the role of ultra-faint satellites is indeed important in this context, as they inform us of star formation in isolated environments as well as the accretion history of the Galaxy.

Here we report on Gaia-ESO Survey observations of two of the recently discovered ultra-faint dwarf satellites: Reticulum~2 and Horologium~1.  Reticulum~2 is a mere $\sim$30\,kpc away towards the Large Magellanic Cloud (LMC) with a $M_v = -2.7 \pm 0.1$, and is suspected to be very dark matter-dominated. Indeed, Reticulum~2 is of particular scientific interest given the recently reported detection of dark matter annihilation \citep{geringer_sameth_2015,fermi_lat_2015}. Horologium~1, also located towards the LMC, is more distant at 79\,kpc but given somewhat brighter luminosity of $M_v = -3.4 \pm 0.1$ and a visible giant branch, spectroscopy of candidate red giant branch (RGB) stars is accessible from the VLT in a standard Gaia-ESO Survey observing block. While our primary goal is to
establish the true nature of these faint stellar systems by gauging
the amount dark matter they might contain, we also intend to test the hypothesis that the dwarfs have once been part of the
Magellanic group.

We outline the target selection and the data that we will subsequently analyse in Section \ref{sec:observations}. A detailed description of our analysis in outlined Section \ref{sec:analysis}. We discuss the interpretations of our results in Section \ref{sec:discussion}.

\section{Observations}
\label{sec:observations}

The data were obtained in 0.8\arcsec{} seeing using the FLAMES spectrograph \citep{pasquini_2002} on the 8.2 m Very Large Telescope (VLT/UT2) at Cerro Paranal, Chile. Candidate members of both satellites were targetted using otherwise unallocated Milky Way fibres on February 6 and March 8-10, 2015 as part of the ongoing Gaia-ESO Survey \citep[ESO Programme 188.B-3002][]{gilmore_2012,randich_2013}. 
In the field \texttt{GES\_MW\_033542\_540254} 109 fibers were allocated to science targets, with 25 allocated to Reticulum~2 candidates, and the rest allocated to standard Gaia-ESO Milky Way targets. In the field \texttt{GES\_MW\_025532\_540711}, the total number of allocated fibers was 107, with 18 fibers assigned to Horologium~1 candidates.
The HR10 and HR21 setups were employed, providing high-resolution ($\mathcal{R} \sim 19800$ and $16200$\footnote{Note however that the resolving power and sensitivity of the GIRAFFE instrument has been recently improved due to refocusing, see http://eso.org/sci/publications/announcements/sciann15013.html}) spectra in wavelength regions of 5334-5611\,\AA{} and 8475-8982\,\AA{}, respectively.

\begin{figure}
\includegraphics{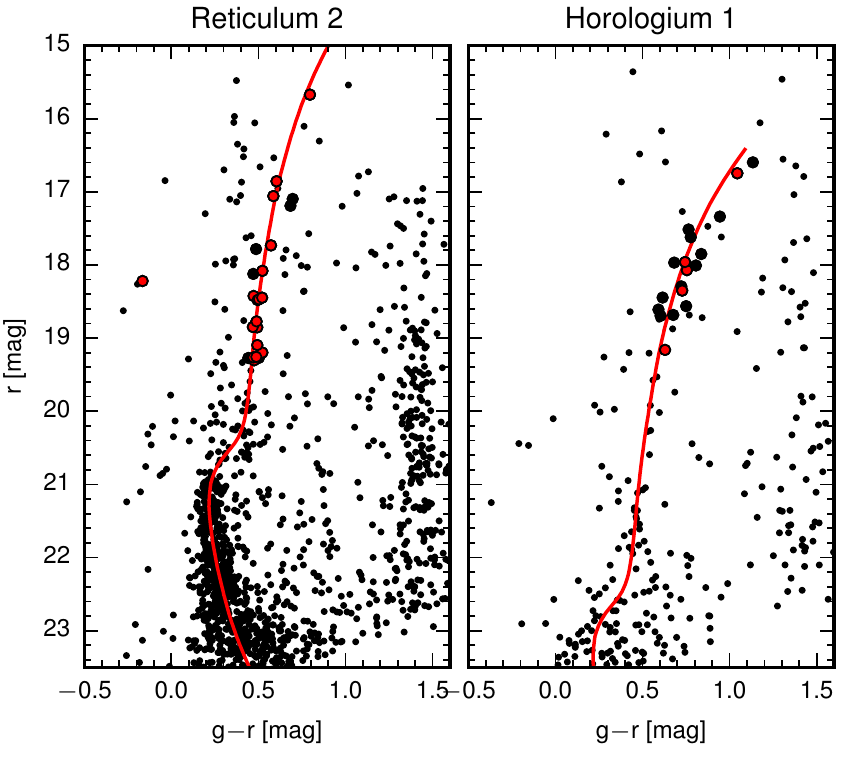}
\caption{Color-magnitude distribution of stars near the centers of Reticulum~2 and Horologium~1 satellites. Large black circles indicate candidates that were observed with VLT/GIRAFFE, and the red symbols are those we later confirmed to be members of each system. Each panel shows a 12.5 Gyr PARSEC isochrone \citet{bressan_2012} with a ${{\rm [Fe/H]} = -2}$ that has been shifted according to the distance modulus from K15. \label{fig0}}
\end{figure}

The candidate satellite members were selected using a broad color-magnitude mask based on the best fitting isochrone and distance modulus from \citet{koposov_2015} (hereafter K15). We also required that the targets were located within 10 - 15$\arcmin$ on the sky from the satellite center. Figure \ref{fig0} shows the color-magnitude distribution of stars near the center of both systems. Candidates that were observed spectroscopically are highlighted, as well as those which we later confirmed to be members.

\begin{figure*}
\plotone{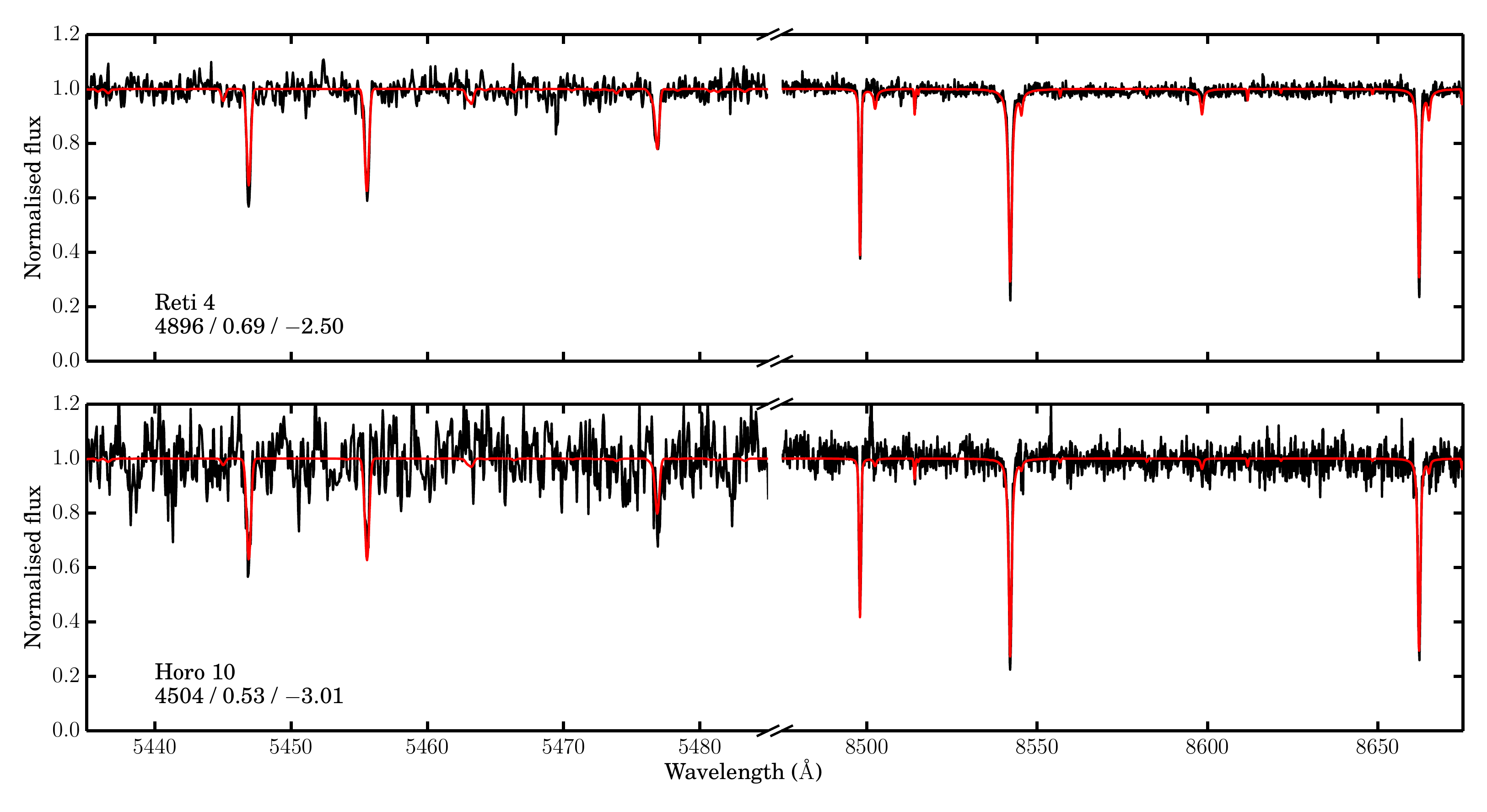}
\caption{An example showing the observed (black) and maximum a posteriori model (red) spectra for a confirmed Reticulum~2 (top) and Horologium~1 (bottom) member. These spectra have the highest S/N ratio of  confirmed members in each system.\label{fig1}}
\end{figure*}

The data were reduced using standard procedures performed for all other Gaia-ESO Survey GIRAFFE observations. This process includes bias correction, flat-fielding, object extraction, sky subtraction, scattered light correction, and wavelength calibration. The spectra are then corrected for barycentric motion and resampled onto a common wavelength scale \citep{lewis_2015}. Generally the mean signal-to-noise (S/N) ratio per pixel is quite low: 5/13 for Reticulum~2 candidates in HR10/HR21 respectively, and 4/10 for the Horologium~1 candidates. However the brightest confirmed members of either satellite (Section \ref{sec:membership-model}) have S/N ratios of 26/62 (Reti 4; HR10/HR21) and 10/28 per pixel (Horo 10; see Figure \ref{fig1}).

\section{Analysis}
\label{sec:analysis}

\subsection{Kinematics}
\label{sec:kinematic-analysis}

The velocity analysis has been performed using standard Gaia-ESO Survey radial velocity (RV) pipeline. Although the full details of this pipeline will be presented in \citet{lewis_2015}, here we outline the principle components. The algorithm is based on a direct pixel-fitting procedure implemented by \citet{koposov11} \citep[see also][]{koleva09,walker_2015a}, and employs the PHOENIX library of model stellar spectra \citep{husser13}. All of the observed spectra are fitted (using a $\chi^2$ metric) by model templates that are interpolated from the spectral library, and the continuum is modelled by a high-degree polynomial. The maximum likelihood point is found by using Nelder-Mead algorithm \citep{nelder_mead_1965}, using several starting points to avoid being trapped in local maxima. The location of the maximum likelihood ($\chi^2$ minimum) and the Hessian of the likelihood surface are used to provide the best-fit velocity, errors and estimates of the stellar atmospheric parameters.

The crucial ingredient in correctly extracting the kinematics of ultra-faint satellites is a proper understanding of the uncertainties of RV measurements \citep[see e.g.,][]{geha_2009,koposov11}. Because the Gaia-ESO Survey has already observed many thousands of stars, some of them repeatedly and with different instrument configurations, our experience with the Survey provide us with a very good understanding of the RV precision achievable. For this paper, however, we focus only on data in just two Gaia-ESO Milky Way fields: those with Horologium~1 and Reticulum~2 candidates.

In addition to the standard data processing steps performed for the Gaia-ESO Survey, there are three subtle points that are important for this study: 

\begin{itemize}

\item \textit{Spectral covariance.} The standard Gaia-ESO reduction pipeline rebins the spectra to a common wavelength mapping with a fixed step size. For HR10 the common wavelength scale extends from 5334-5611\AA{}  and for HR21 the boundaries are 8475-8982\AA{}, both with a spacing of 0.05\AA{}. While convenient for some analyses, the rebinning procedure introduces correlated noise/covariance in the spectra and reduces the effective information content of the spectra. For example, the rebinned HR21 spectra has 10141 pixels, the rebinned HR10 spectra has 5541 pixels, while the original spectra are just 4096 pixels. We can account for this correlation by modelling the spectra with the full covariance matrix, or approximate it. Our tests found that if we fit the spectra using the full covariance matrix of the data and the posterior/likelihood is properly behaved (e.g., uni-modal and close to a Gaussian), then the effect of pixel covariance is equivalent to scaling the  errors by a fixed constant: 1.5 for HR10, 2.0 for HR21. These numbers are approximately equal to the ratio of rebinned and original pixels. We adopt this scaling throughout the rest of our analysis. See the end of this section for the verification of the results.

\item \textit{Systematic error floor.} It is well-known that although the formal RV precision derived from cross-correlation or pixel-fitting methods can be almost arbitrarily small for sufficiently high S/N spectra, the actual precision achievable with most spectrographs is generally limited by systematic effects. This includes instrument flextures, uncertainties in the wavelength calibrations, Line Spread Function (LSF) variation/asymmetry and template mismatches. This systematic component has to be included in the total error budget. We have found this systematic error to be around 300\,m s$^{-1}$ from large numbers of Gaia-ESO Milky Way spectra. It is important to note that this systematic component is not expected to be present when comparing RVs obtained from spectra using the same setup in sequential exposures, but it becomes important when comparing RVs from different nights, or between HR10 and HR21 setups. We include this systematic error floor in suitable comparisons hereafter.

\item \textit{RV offset between HR10 and HR21.} Over the course of the Gaia-ESO Survey, it has been discovered that there is a small systematic offset of 400\,m~s$^{-1}$ between the RVs measured in the HR10 and HR21 setups. The cause of this offset is not well-established yet. This correction was applied to the radial velocities (the HR21 velocities have been shifted by $-400$\,m~s$^{-1}$).

\end{itemize}

\begin{figure}
\includegraphics{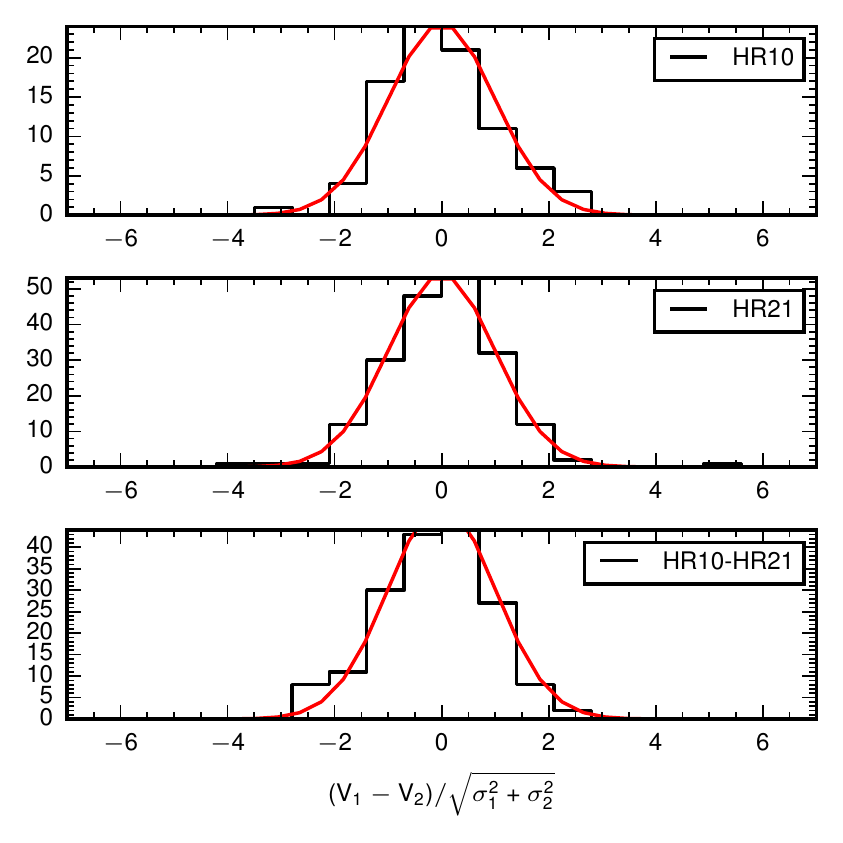}
\caption{The results of radial velocity precision tests done using repeated observations. The three panels show the distribution of the RV difference normalised by their errors as measured in the HR10 (top panel), HR21 (middle panel) and HR10-HR21 configurations (bottom panel). These tests demonstrate that our error model is correct, as the distributions very closely resemble a normal distribution with zero-mean and unit variance shown by red curves on the panels.}
\label{fig:rvrepeat}
\end{figure}

After applying the aforementioned corrections, we can confirm whether the RVs measured in repeated exposures match within the precision quoted by our error bars. To test this we have collated all the spectra in the \texttt{GES\_MW\_033542\_540254} and \texttt{GES\_MW\_025532\_540711} fields (i.e., including both standard Gaia-ESO targets and possible satellite member stars). The top panel of Figure~\ref{fig:rvrepeat} shows the distribution of velocity differences ($V_1-V_2$) scaled by the RV error ($\sqrt{\sigma(V_1)^2+\sigma(V_2)^2}$) for repeated HR10 exposures. The middle panel of the figure shows the same for the HR21 setup. The bottom panel shows the distribution of normalised velocity differences for HR10 exposures versus HR21. In all the panels the red curve shows a standard normal distribution with zero mean and  unit variance. In all cases the distributions are indeed well described by Gaussians, confirming that our error model provides a correct description of the velocity uncertainties.

The final radial velocities for all the observed Reticulum~2 and Horologium~1 candidate members are provided in Table~\ref{tbl2} and refer to the weighted means of the HR10 and HR21 measurements and take into account all the error-terms mentioned above.

\subsection{Chemistry}
\label{sec:chemical-analysis}
We used a generative model to infer the stellar parameters for all stars. The model is described as follows. For a given set of stellar parameters ${\omega = \{T_{\rm eff},\log{g},{\rm [Fe/H]},[\alpha/{\rm Fe}]\}}$ we first produce a flux-normalised synthetic spectrum $S(\lambda,\omega)$ at wavelengths $\lambda$ by interpolating spectra from a surrounding grid. The synthetic spectra were calculated as per the AMBRE grid \citep[see][for details]{de_laverny_2012}. This high-resolution ($\mathcal{R} > 300,000$) grid was synthesised specifically for the Gaia-ESO Survey using Turbospectrum \citep{alvarez_1998,plez_2012}, the MARCS \citep{marcs_2008} model atmospheres and the Gaia-ESO Survey line list \citep[][V5 for atoms and molecules]{ruffoni_2014,heiter_2015}. The grid includes effective temperatures from 3000 K to 8000 K, and surface gravities from $\log{g} = 0$ to 5. Metallicities extend from as low as ${\rm [Fe/H]} = -5$ with 1\,dex steps until ${\rm [Fe/H]} = -3$ and 0.25\,dex steps thereafter, extending past solar metallicity. In the metallicity range applicable for this study, $[\alpha/{\rm Fe}]$ ratios vary between $0.0$ and $+0.8$ at steps of 0.2\,dex. We redshift our interpolated spectrum by velocity $V$ such that the normalised synthetic flux at an observed point $\lambda$ is given by $S\left(\lambda \left[1 + \frac{V}{c}\right],\omega\right)$, where $c$ is the speed of light. The observed continuum is modelled as a low-order polynomial with coefficients $b_j$ that enters multiplicatively:
\begin{equation}
    M(\lambda,\omega,v,\{b\}) = \sum_{j=0}^{N-1}b_{channel,j}\lambda^{j} \times S\left(\lambda\left[1 + \frac{V}{c}\right],\omega\right)
\end{equation}

The continuum in each observed channel (HR10 and HR21) are modelled separately. In practice we found a first-order polynomial to sufficiently represent the continuum in each channel. Lastly, we convolve the model spectrum with a Gaussian LSF (with free parameter $\mathcal{R}$) to match the resolving power in each channel, and resample the model spectrum to the observed pixels $\{\lambda\}$. Although the spectral resolution $\mathcal{R}$ in each channel is reasonably well-known, recent refocusing of the GIRAFFE spectrograph has improved the quoted spectral resolution. For this reason we chose to include the spectral resolution $\mathcal{R}$ as a nuisance parameter with reasonable priors and marginalise them away. The prior on spectral resolution was uniformly distributed to within $\pm$30\% of $\mathcal{R} = $16200 and 19800 for the HR10 and HR21 setups respectively. After convolution with the LSF, binning to the observed pixels $\{\lambda\}$ and assuming Gaussian error $\sigma_i$, the probability distribution $p\left(F_i|\lambda_i,\sigma_i,\omega,V,\{b\},\{\mathcal{R}\}\right)$ for the observed spectral flux $F_i$ is:
\begin{equation}
p\left(F_{i}|\lambda_i,\sigma_i,\omega,V,\{b\},\{\mathcal{R}\}\right) = \frac{1}{\sqrt{2\pi\sigma_{i}^{2}}}\exp{\left(-\frac{\left[F_i - M_i\right]^2}{2\sigma_{i}^{2}}\right)}.
\end{equation}

Under the implied assumption that the data are independently drawn, the likelihood of observing the data $D$, given our model, is found by the product of individual probabilities:

\begin{equation}
    \mathcal{L} = \prod_{i=1}^{N}\,p\left(F_i|\lambda_i,\sigma_i,\omega,V,\{b\},\{\mathcal{R}\}\right)
\end{equation}

\noindent{}and the probability $\mathcal{P}$ of observing the data is proportional up to a constant such that:

\begin{eqnarray}
\mathcal{P}     & \propto   & \mathcal{L}\left(D|\theta\right) \times \mathcal{P}r\left(\theta\right) \nonumber \\
\ln\mathcal{P}  &   =       & \ln\mathcal{L}\left(D|\theta\right) + \ln\mathcal{P}r\left(\theta\right) 
\end{eqnarray}

\noindent{}where $\mathcal{P}r(\theta)$ is the prior probability on the model parameters $\theta$. 

In practice we found that the spectral range of our data were not particularly informative of the effective temperature $T_{\rm eff}$ for very metal-poor stars. For these stars the data were prone to favour unphysically cool super giant stars of extremely low metallicity. We did not find the same effect for more metal-rich stars, where there are sufficient neutral and ionising transitions present to accurately constrain the stellar parameters. Given most of our candidates are indeed metal-poor, we found it prudent to fix the effective temperature using the DES photometry and a colour-temperature relation\footnote{The colour-temperature relation was defined by fitting the relation between DES g$-$r colors and effective temperature from SEGUE \citep{lee_2008}.}. We found this had no significant impact on our posteriors for the foreground dwarfs -- where the spectra are indeed informative of effective temperature -- and ultimately did not substantially alter our inferred metallicity dispersion for either satellite. Only the mean satellite metallicities were affected. The uncertainties in effective temperature listed in Table \ref{tbl2} were calculated by propagating the DES photometric uncertainties with the intrinsic uncertainty in the colour-temperature relation. Thus, our forward model is subject to our photometric temperatures and has only 10 parameters: $\log{g}$, [Fe/H], [$\alpha$/Fe], $V$, the resolving powers $\mathcal{R}_{\rm HR10}$ and $\mathcal{R}_{\rm HR21}$, as well as two continuum coefficients in each channel.

The initial model parameters $V$ and $\omega$ (modulo $T_{\rm eff}$) were estimated by performing a coarse normalisation of the data and cross-correlating it against the \citet{de_laverny_2012} grid. Although we have fixed $T_{\rm eff}$ and previously determined $V$ (see Section \ref{sec:kinematic-analysis}) we still carried out the cross-correlation to yield a reliable initial estimate of $\log{g}$, [Fe/H] and [$\alpha$/Fe]. We also used the synthetic flux at the grid point with the peak cross-correlation coefficient to subsequently estimate the normalisation coefficients $\{b\}$. 
We numerically optimised the negative log-probability $-\ln{\mathcal{P}}$ from the initial point using the Nelder-Mead algorithm \citep{nelder_mead_1965}. Following optimisation, we sampled the resulting posterior using the affine-invariant Markov Chain Monte Carlo (MCMC) sampler introduced by \citet{goodman_weare_2010} and implemented by \citet{foreman_mackey_2013}. In all cases at least 200 walkers were used to explore the parameter space for more than 2500 steps ($\geq5\times10^5$ probability evaluations) to burn-in the sampler. These probability calls were discarded and the chains were reset before production sampling began. We tested our MCMC analyses for convergence by examining the auto-correlation times (e.g., ensuring high effective sample numbers per parameter) and the mean acceptance fractions over time. We also re-ran a subset of our analyses with many more evaluations (for both burn-in and production), verifying that there was no change to the resulting posteriors.

\begin{figure*}[t!]
\plotone{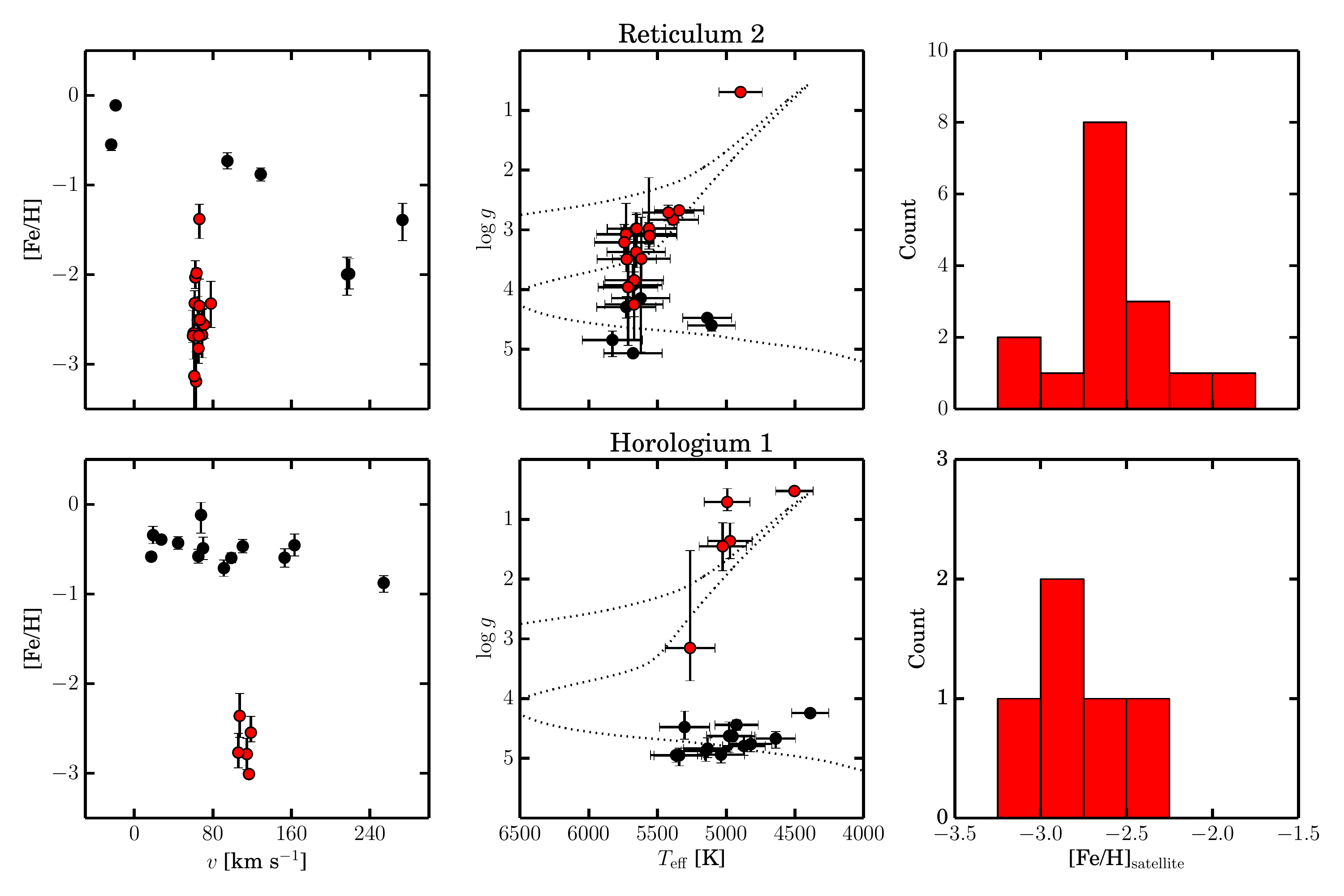}
\caption{Inferred radial velocities and stellar parameters for all Reticulum~2 (top) and Horologium~1 (bottom) candidates. Confirmed members (marked in red) cleanly separate from foreground contaminants in $v-{\rm [Fe/H]}$ space (left). Photometric effective temperatures (see text) and spectroscopically-derived surface gravities are shown in the middle panel, overlaid upon a 12 Gyr PARSEC isochrone \citep{bressan_2012} with a metallicity of ${\rm [Fe/H]} = -2$. Histograms of the maximum a posteriori metallicities of confirmed members are shown in the right-hand panels.\label{fig:stellar_params}}
\end{figure*}

We list the photometric effective temperatures and other inferred stellar parameters (given the model and effective temperature) in Table \ref{tbl2}. The middle panels of Figure \ref{fig:stellar_params} show the inferred surface gravities for all candidates. Our confirmed members agree well with the metal-poor isochrone shown. This is particularly true for the higher-quality Reticulum~2 data, with the possible exception of the metal-poor giant Reti~4, where the $\log{g}$ seems quite low. Although this star has the highest S/N ratio in our sample, metal-poor super giant stars are very challenging to model from an astrophysical perspective. Nevertheless, the marginalised posterior metallicity distribution for Reti~4 agrees excellently with stars further down the giant branch of lower S/N ratios. Table~\ref{tbl2} also lists the reduced $\chi^2$ values of our spectral fits. In general the values are quite close to 1. However due to lower S/N and imperfect sky-subtraction in the Horologium~1 data, the $\chi^2$/d.o.f. values for those candidates are slightly higher.

Our radial velocity determination for Reti~22 confirms it as a horizontal branch member of Reticulum~2 (see Section \ref{sec:membership-model}). However, the photometric temperature estimate of 8468$^{+306}_{-295}$\,K prohibited us from inferring other stellar parameters for this star, as it is hotter than the boundary (8000\,K) of the spectral grid. 

Two members of Reticulum~2 have maximum a posteriori (MAP) $\log{g}$ values that are consistent with being a dwarf (Figure \ref{fig:stellar_params}). This is inconsistent with the photometry, since the main-sequence is too faint for us to target with standard exposure times for Gaia-ESO Survey Milky Way fields. However, the negative uncertainties on $\log{g}$ for these two Reticulum~2 members are considerably large, making them deviate from the giant branch by only 1 to 1.5$\sigma$. As a test we constrained the prior on surface gravity to be uninformative between $\log{g} \in [-0.5, 4.0]$, forcing the star to be a giant/sub-giant, but we found no statistically significant difference in the marginalized posterior metallicity distribution.

Similarly while all confirmed Horologium~1 stars are giants, and the foreground contaminants are clearly dwarfs, one star (Horo~17) has a very low S/N and consequently has an extremely large negative uncertainty in $\log{g}$. While the posterior demonstrates the star is not a dwarf, we cannot place its precise location on the giant branch.

Our inferred [$\alpha$/Fe] abundance ratios are informative, even with their large uncertainties. Unsurprisingly, we found the [$\alpha$/Fe] ratio to be strongly correlated with other stellar parameters, particularly [Fe/H]. We find the foreground contaminants to largely follow the well-studied Milky Way trend in ${\rm [Fe/H]}-[\alpha/{\rm Fe}]$. All the confirmed members in Reticulum~2 and Horologium~1 appear to have at least $[\alpha/{\rm Fe}] \gtrsim +0.2$, with an point estimate (assumed $\delta$-function distribution) of $[\alpha/{\rm Fe}] \approx +0.4$ in Reticulum~2 and $\approx{}+0.3$ in Horologium~1.

\begin{deluxetable}{lcc}
\tablecaption{Summary properties of Reticulum~2 and Horologium~1\label{tbl-1}}
\tablewidth{0pt}
\tablehead{\colhead{} & \colhead{Reticulum~2} & \colhead{Horologium~1}} 
\startdata
$\alpha$ (J2000) [deg]  &   53.9256 &   43.8820 \\
$\delta$ (J2000) [deg]  & $-$54.0592 & $-$54.1188 \\
Distance [kpc]          &        30 &        79 \\
$M_V$                   &  $-2.7 \pm 0.1$ & $-3.4 \pm 0.1$ \\
Ellipticity             &  0.59$^{+0.02}_{-0.03}$ & $<0.28$ \\
$r_{1/2}$ [arcmin]      &  3.64$^{+0.21}_{-0.12}$ & $1.31^{+0.19}_{-0.14}$ \\
$r_{1/2}$ [pc]          &  32$^{+1.9}_{-1.1}$ & $30^{+4.4}_{-3.3}$ \\
\tableline
$V_{hel}$ [km s$^{-1}$] & 64.7$^{+1.3}_{-0.8}$&  112.8$^{+2.5}_{-2.6}$\\

$\sigma\left(V\right)$ [km s$^{-1}$] & $3.22^{+1.64}_{-0.49}$ & $4.9^{+2.8}_{-0.9}$ \\

Mass($<r_{1/2}$) [$M_\odot$] & $2.35^{+4.71}_{-0.13}\times 10^5$ & $5.5^{+11.3}_{-1.0} \times 10^5$  \\

$M/L_V [M_\odot/L_\odot]$ & $479^{+904}_{-51}$ & $570^{+1154}_{-112}$ \\

${\rm [Fe/H]}$ & $-2.46^{+0.09}_{-0.1}$ & $-2.76^{+0.1}_{-0.1}$ \\

$\sigma\left({\rm [Fe/H]}\right)$ [dex] & $0.29^{+0.13}_{-0.05}$ & $0.17^{+0.2}_{-0.03}$\\
$[\alpha/{\rm Fe}]$ & $0.40 \pm 0.04$ & $0.30 \pm 0.07$ 
\enddata
\tablecomments{Properties above the horizontal separator were adopted from \citet{koposov_2015}. Those below the separator were determined in this study.}
\end{deluxetable}

\begin{deluxetable*}{lcccrrrrrcc}
\tablecaption{Positions, velocities, stellar parameters and membership for Reticulum~2 and Horologium~1 candidates\label{tbl2}}
\tablewidth{0pt}
\tablehead{
    \colhead{Object} &
    \colhead{$\alpha$ (J2000)} &
    \colhead{$\delta$ (J2000)} &
    \colhead{$g$} &
    \colhead{$V_{hel}$} &
    \colhead{$T_{\rm eff}$} &
    \colhead{$\log{g}$} &
    \colhead{[Fe/H]} &
    \colhead{[$\alpha$/Fe]} &
    \colhead{$\chi^2_{red}$} &
    \colhead{Member?} \\
 & \colhead{[deg]} & \colhead{[deg]} & \colhead{[mag]} & \colhead{[km s$^{-1}$]} & \colhead{[K]}}
\startdata
\\
\multicolumn{10}{c}{\textbf{Reticulum~2}} \\ \tableline \\
Reti~0   & 53.9424   & $-54.1260$  & 19.27 & $218.5 \pm 1.1$ & $5680^{+217}_{-209}$    & $3.93^{+0.35}_{-0.33}$ & $-1.99^{+0.17}_{-0.17}$ &  $0.55^{+0.17}_{-0.19}$ & 1.11 &     \\
Reti~1   & 53.8133   & $-54.1452$  & 19.78 & $ 78.9 \pm 1.8$ & $5714^{+221}_{-213}$    & $3.96^{+0.80}_{-0.98}$ & $-2.32^{+0.27}_{-0.25}$ &  $0.61^{+0.14}_{-0.22}$ & 0.96 & Yes \\
Reti~2   & 53.8072   & $-54.0824$  & 19.74 & $ 60.0 \pm 2.1$ & $5729^{+219}_{-211}$    & $3.07^{+0.52}_{-0.43}$ & $-2.65^{+0.29}_{-0.32}$ &  $0.45^{+0.24}_{-0.28}$ & 0.95 & Yes \\
Reti~3   & 53.9045   & $-54.0670$  & 18.60 & $ 65.6 \pm 0.9$ & $5558^{+204}_{-197}$    & $3.10^{+0.17}_{-0.18}$ & $-2.82^{+0.17}_{-0.12}$ &  $0.65^{+0.11}_{-0.17}$ & 1.00 & Yes \\
Reti~4   & 53.8494   & $-54.0687$  & 16.47 & $ 66.3 \pm 0.2$ & $4896^{+160}_{-155}$    & $0.69^{+0.02}_{-0.02}$ & $-2.50^{+0.02}_{-0.02}$ &  $0.23^{+0.01}_{-0.01}$ & 1.43 & Yes \\
Reti~5   & 53.8374   & $-54.0633$  & 18.97 & $ 69.1 \pm 1.0$ & $5655^{+216}_{-208}$    & $3.37^{+0.63}_{-0.26}$ & $-2.54^{+0.18}_{-0.16}$ &  $0.59^{+0.14}_{-0.22}$ & 1.03 & Yes \\
Reti~6   & 53.7260   & $-54.0994$  & 18.97 & $ 70.8 \pm 1.1$ & $5617^{+214}_{-206}$    & $3.48^{+0.69}_{-0.36}$ & $-2.56^{+0.15}_{-0.25}$ &  $0.14^{+0.20}_{-0.11}$ & 0.91 & Yes \\
Reti~7   & 53.7399   & $-54.0920$  & 18.97 & $ 61.9 \pm 0.8$ & $5564^{+207}_{-199}$    & $3.07^{+0.19}_{-0.17}$ & $-2.03^{+0.12}_{-0.19}$ &  $0.39^{+0.17}_{-0.15}$ & 1.05 & Yes \\
Reti~8   & 53.7605   & $-54.0650$  & 19.26 & $ 65.4 \pm 1.8$ & $5669^{+217}_{-209}$    & $3.84^{+0.89}_{-0.61}$ & $-2.51^{+0.27}_{-0.26}$ &  $0.29^{+0.30}_{-0.22}$ & 0.91 & Yes \\
Reti~9   & 53.8209   & $-54.0675$  & 19.74 & $ 62.9 \pm 3.7$ & $5671^{+216}_{-208}$    & $4.25^{+0.55}_{-0.60}$ & $-3.13^{+0.39}_{-0.45}$ &  $0.22^{+0.30}_{-0.20}$ & 1.44 & Yes \\
Reti~10  & 53.8540   & $-54.0418$  & 19.72 & $ 65.6 \pm 1.2$ & $5562^{+207}_{-200}$    & $2.98^{+0.85}_{-0.34}$ & $-1.38^{+0.22}_{-0.16}$ &  $0.42^{+0.20}_{-0.22}$ & 0.98 & Yes? \\
Reti~11  & 53.7986   & $-54.0560$  & 19.35 & $ 68.2 \pm 1.7$ & $5651^{+217}_{-209}$    & $2.98^{+0.24}_{-0.22}$ & $-2.67^{+0.25}_{-0.28}$ &  $0.39^{+0.26}_{-0.26}$ & 1.17 & Yes \\
Reti~12  & 53.7896   & $-54.0416$  & 18.27 & $ 94.7 \pm 0.4$ & $5678^{+216}_{-208}$    & $5.07^{+0.01}_{-0.01}$ & $-0.73^{+0.09}_{-0.09}$ &  $0.31^{+0.07}_{-0.06}$ & 1.02 &     \\
Reti~13  & 54.0980   & $-54.0885$  & 19.77 & $220.3 \pm 2.0$ & $5622^{+207}_{-200}$    & $4.15^{+0.77}_{-0.90}$ & $-2.00^{+0.23}_{-0.19}$ &  $0.53^{+0.17}_{-0.25}$ & 1.04 &     \\
Reti~14  & 54.0074   & $-54.0681$  & 19.59 & $ 63.4 \pm 1.7$ & $5655^{+218}_{-210}$    & $2.98^{+0.27}_{-0.25}$ & $-3.19^{+0.31}_{-0.33}$ &  $0.46^{+0.23}_{-0.27}$ & 1.14 & Yes \\
Reti~15  & 53.9502   & $-54.0638$  & 18.30 & $ 63.5 \pm 0.5$ & $5421^{+191}_{-185}$    & $2.71^{+0.12}_{-0.14}$ & $-1.98^{+0.06}_{-0.05}$ &  $0.47^{+0.06}_{-0.07}$ & 1.11 & Yes \\
Reti~16  & 53.9582   & $-54.0559$  & 19.71 & $292.1 \pm 1.3$ & $5828^{+223}_{-215}$    & $4.84^{+0.15}_{-0.28}$ & $-1.39^{+0.23}_{-0.18}$ &  $0.54^{+0.17}_{-0.23}$ & 1.06 &     \\
Reti~17  & 53.9845   & $-54.0545$  & 18.90 & $ 65.9 \pm 1.2$ & $5724^{+220}_{-211}$    & $3.49^{+0.26}_{-0.21}$ & $-2.68^{+0.21}_{-0.30}$ &  $0.22^{+0.29}_{-0.17}$ & 1.09 & Yes \\
Reti~18  & 54.0323   & $-54.0432$  & 17.46 & $ 61.4 \pm 0.4$ & $5343^{+183}_{-177}$    & $2.67^{+0.07}_{-0.06}$ & $-2.32^{+0.04}_{-0.14}$ &  $0.42^{+0.16}_{-0.02}$ & 1.05 & Yes \\
Reti~19  & 53.9923   & $-54.0346$  & 19.32 & $ 65.0 \pm 1.4$ & $5741^{+221}_{-212}$    & $3.21^{+0.30}_{-0.22}$ & $-2.35^{+0.23}_{-0.30}$ &  $0.22^{+0.29}_{-0.22}$ & 1.06 & Yes \\
Reti~20  & 54.0163   & $-54.0073$  & 17.79 & $-23.6 \pm 0.3$ & $5107^{+177}_{-171}$    & $4.60^{+0.07}_{-0.09}$ & $-0.55^{+0.07}_{-0.05}$ &  $0.19^{+0.01}_{-0.06}$ & 1.06 &     \\
Reti~21  & 54.0952   & $-53.9987$  & 18.59 & $128.9 \pm 0.5$ & $5727^{+219}_{-211}$    & $4.29^{+0.17}_{-0.19}$ & $-0.88^{+0.07}_{-0.07}$ & $-0.01^{+0.06}_{-0.06}$ & 0.97 &     \\
Reti~22  & 54.0779   & $-53.9625$  & 18.05 & $ 61.6 \pm 2.6$ & $8468^{+306}_{-295}$    & \nodata                & \nodata                 & \nodata                 & \nodata & Yes \\
Reti~23  & 53.8798   & $-54.0300$  & 17.64 & $ 59.6 \pm 0.5$ & $5386^{+187}_{-181}$    & $2.83^{+0.09}_{-0.09}$ & $-2.68^{+0.07}_{-0.28}$ &  $0.47^{+0.30}_{-0.07}$ & 1.16 & Yes \\
Reti~24  & 53.9127   & $-53.9323$  & 17.88 & $-19.2 \pm 0.3$ & $5138^{+181}_{-175}$    & $4.48^{+0.06}_{-0.05}$ & $-0.11^{+0.03}_{-0.03}$ &  $0.09^{+0.02}_{-0.02}$ & 1.00 &  \\
\tableline\\
\multicolumn{10}{c}{\textbf{Horologium~1}} \\ \tableline \\
Horo~0   & 43.9692   & $-54.3117$  & 18.81 & $110.6 \pm 0.4$ & $5135^{+179}_{-173}$    & $4.84^{+0.18}_{-0.15}$ & $-0.47^{+0.07}_{-0.08}$ &  $0.30^{+0.06}_{-0.06}$ & 2.01 &     \\
Horo~1   & 44.0306   & $-54.2768$  & 18.28 & $254.2 \pm 0.2$ & $4640^{+146}_{-141}$    & $4.67^{+0.12}_{-0.16}$ & $-0.88^{+0.10}_{-0.08}$ &  $0.29^{+0.04}_{-0.04}$ & 1.58 &     \\
Horo~2   & 43.8105   & $-54.2267$  & 19.20 & $152.6 \pm 0.8$ & $5365^{+189}_{-182}$    & $4.95^{+0.05}_{-0.12}$ & $-0.59^{+0.10}_{-0.10}$ &  $0.38^{+0.10}_{-0.11}$ & 2.23 &     \\
Horo~3   & 43.6503   & $-54.1802$  & 17.73 & $ 27.5 \pm 0.1$ & $4389^{+137}_{-132}$    & $4.24^{+0.05}_{-0.02}$ & $-0.39^{+0.02}_{-0.03}$ &  $0.21^{+0.03}_{-0.01}$ & 2.03 &     \\
Horo~5   & 44.1126   & $-54.2174$  & 17.79 & $ 64.8 \pm 0.5$ & $5038^{+173}_{-168}$    & $4.94^{+0.07}_{-0.14}$ & $-0.58^{+0.08}_{-0.08}$ &  $0.29^{+0.07}_{-0.07}$ & 1.58 &     \\
Horo~6   & 44.1567   & $-54.1941$  & 19.06 & $ 91.2 \pm 0.9$ & $4871^{+158}_{-153}$    & $4.79^{+0.16}_{-0.15}$ & $-0.71^{+0.09}_{-0.09}$ &  $0.38^{+0.05}_{-0.08}$ & 2.22 &     \\
Horo~7   & 44.0076   & $-54.1986$  & 19.30 & $163.1 \pm 1.0$ & $5148^{+182}_{-176}$    & $4.88^{+0.10}_{-0.17}$ & $-0.46^{+0.12}_{-0.13}$ &  $0.39^{+0.12}_{-0.10}$ & 1.53 &     \\
Horo~9   & 43.9179   & $-54.1353$  & 18.71 & $118.5 \pm 0.5$ & $4993^{+168}_{-163}$    & $0.71^{+0.23}_{-0.15}$ & $-2.55^{+0.11}_{-0.18}$ &  $0.35^{+0.17}_{-0.14}$ & 1.97 & Yes \\
Horo~10  & 43.8967   & $-54.1122$  & 19.31 & $116.6 \pm 0.1$ & $4504^{+138}_{-134}$    & $0.53^{+0.04}_{-0.02}$ & $-3.01^{+0.02}_{-0.03}$ &  $0.36^{+0.03}_{-0.03}$ & 1.83 & Yes \\
Horo~11  & 43.9699   & $-54.0877$  & 18.83 & $114.6 \pm 0.7$ & $4972^{+163}_{-158}$    & $1.36^{+0.30}_{-0.29}$ & $-2.79^{+0.17}_{-0.17}$ &  $0.35^{+0.19}_{-0.18}$ & 2.24 & Yes \\
Horo~15  & 43.8912   & $-54.0939$  & 19.08 & $105.6 \pm 1.0$ & $5026^{+174}_{-168}$    & $1.45^{+0.39}_{-0.41}$ & $-2.77^{+0.17}_{-0.22}$ &  $0.15^{+0.21}_{-0.11}$ & 2.02 & Yes \\
Horo~17  & 43.8719   & $-54.0727$  & 18.65 & $108.1 \pm 1.9$ & $5263^{+184}_{-178}$    & $3.15^{+1.63}_{-0.55}$ & $-2.36^{+0.24}_{-0.25}$ &  $0.18^{+0.27}_{-0.17}$ & 2.03 & Yes \\
Horo~18  & 43.8497   & $-54.0445$  & 18.40 & $ 17.1 \pm 0.2$ & $4925^{+160}_{-155}$    & $4.44^{+0.08}_{-0.09}$ & $-0.58^{+0.05}_{-0.05}$ &  $0.16^{+0.05}_{-0.04}$ & 4.25 &     \\
Horo~19  & 43.8867   & $-53.9968$  & 19.79 & $ 18.7 \pm 0.6$ & $5303^{+185}_{-179}$    & $4.48^{+0.27}_{-0.20}$ & $-0.34^{+0.10}_{-0.10}$ &  $0.02^{+0.09}_{-0.10}$ & 1.57 &     \\
Horo~20  & 43.6278   & $-54.0217$  & 18.28 & $ 98.5 \pm 0.2$ & $4954^{+164}_{-159}$    & $4.63^{+0.11}_{-0.10}$ & $-0.60^{+0.05}_{-0.06}$ &  $0.08^{+0.04}_{-0.05}$ & 3.00 &     \\
Horo~21  & 43.6917   & $-53.9571$  & 19.36 & $ 67.4 \pm 1.1$ & $5342^{+186}_{-180}$    & $4.95^{+0.12}_{-0.17}$ & $-0.12^{+0.20}_{-0.14}$ & $-0.12^{+0.12}_{-0.11}$ & 2.96 &     \\
Horo~22  & 43.9334   & $-53.9473$  & 19.02 & $ 70.0 \pm 0.6$ & $4979^{+167}_{-161}$    & $4.63^{+0.23}_{-0.28}$ & $-0.49^{+0.13}_{-0.12}$ &  $0.06^{+0.12}_{-0.10}$ & 1.99 &     \\
Horo~23  & 43.8368   & $-53.9240$  & 18.69 & $ 44.2 \pm 0.3$ & $4821^{+155}_{-150}$    & $4.76^{+0.12}_{-0.13}$ & $-0.43^{+0.07}_{-0.07}$ &  $0.25^{+0.06}_{-0.07}$ & 1.73 &
\enddata
\end{deluxetable*}

\subsection{Joint modeling of satellites kinematics and chemistry}
\label{sec:membership-model}

\begin{figure*}
\includegraphics{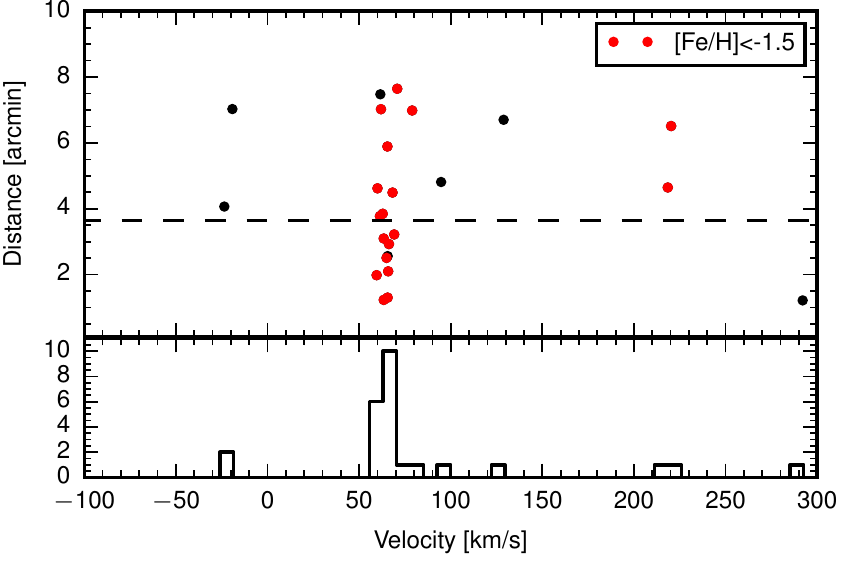}
\includegraphics{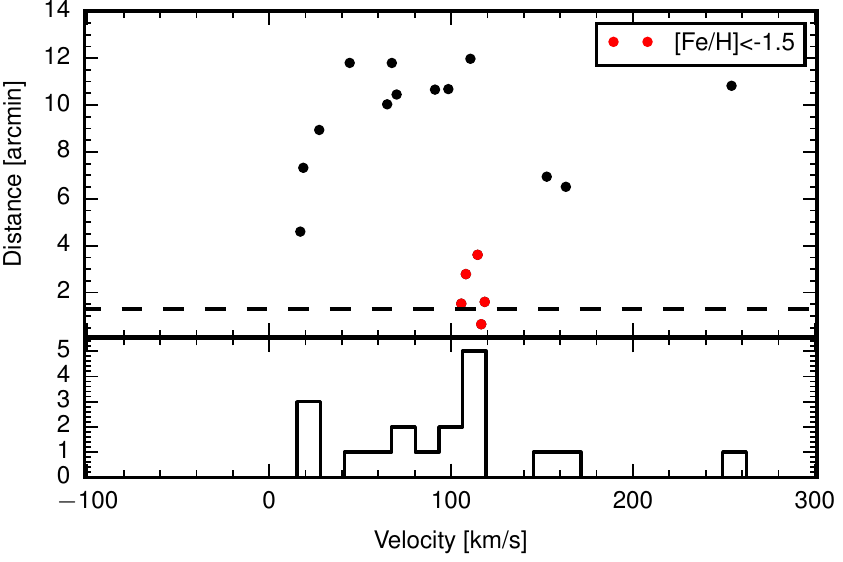}

\caption{The measured RV and spatial distance from the satellite centroid for Reticulum~2 (left) and Horologium~1 (right). The bottom panels show the RV distributions. The dashed line on top panels indicates the half-light radius as measured by K15. Red circles are the stars with ${\rm [Fe/H]} < -1.5$. In the case of Reticulum~2 the peak in the radial velocities at V$\sim 65$\,km s$^{-1}$ due to the satellite stars is obvious. In the case of Horogolium~1 the peak at V$\sim 100$\,km s$^{-1}$ is less prominent but still significant, because all the stars in that peak are within $\sim$ twice the half-light radius and have low metallicity as opposed to high metallicity background stars located at larger distances from the satellite center.}
\label{fig:rv_dist}
\end{figure*}

Having measured the chemical abundances and radial velocities of individual stars in two satellites the next step is combining all the available information in order to obtain the most reliable inference on the average velocity, metallicity and their dispersions, while properly accounting for any potential foreground contamination.
Figure~\ref{fig:rv_dist} shows the radial velocity of stars versus the distance to the center of two satellites. It clearly illustrates that although the velocity signal due to the satellites is quite prominent, the contamination -- albeit minor -- still has to be taken into account. Thus, in order to describe the velocity distribution of each satellite we adopt the following set of mixture models \citep[see e.g.][for similar approach]{walker09,koposov11}.

\begin{eqnarray}
P(V, \psi | \phi) &  = & f \, P_{sat}(\psi)\, {\mathcal N}(V|V_0,\sigma) + \nonumber \\
& & (1  - f)\, P_{bg}(\psi)\, {\mathcal N}(V|V_{bg},\sigma_{bg}) 
\label{eq:rvmodel}
\end{eqnarray} 

Where $V$ is the heliocentric velocity, $\mathcal{N}$ is a Gaussian distribution, $f$ is the fraction of objects belonging to the satellite, $\phi$ is the shorthand notation for the parameters of the model  and $\psi$ are ancillary variables that help us identify members (such as metallicity and/or distance from the center of the object). The key assumption is that the RV distribution for each of the satellites is Gaussian and RVs of background/foreground stars are also Gaussian distributed (a reasonable assumption given very small number of such stars).

Since each RV measurement comes with an error bar, the actual likelihood of each RV point $V_i$ and error $\sigma_i$ is the convolution of the model from Eq.~\ref{eq:rvmodel} with the Gaussian error:  $P(D|\phi) \propto \int P(V| \phi)\, {\mathcal N}(V|V_i,\sigma_i)\,dv$. Given that the underlying velocity models $P(V| \phi)$ are Gaussians themselves, the integral is trivial to analytically compute.

The ancilliary parameters $\psi$ serve the purpose of helping to separate the satellite members from the background stars. For Reticulum~2 we use $\psi ={\rm [Fe/H]}$ and model the joint distribution of metallicity and RV. We assume that the metallicities of both the background and the object are Gaussian-distributed (with different means and variances):
$P_{sat}({\rm [Fe/H]})={\mathcal N}({\rm [Fe/H]}|{\rm [Fe/H]}_{sat},\sigma_{{\rm [Fe/H]},{sat}})$, $P_{bg}({\rm [Fe/H]})={\mathcal N}({\rm [Fe/H]}|{\rm [Fe/H]}_{bg},\sigma_{{\rm [Fe/H]},{bg}})$.
For Horologium~1 we have fewer potential members, so we require more information than RV and metallicity. Therefore we model the joint distribution of RV, metallicity, and distance from the center of the satellite: $\psi=\{r,{\rm [Fe/H]}\}$.
The metallicities are modelled as Gaussian distributions, while an exponential density model with the morphological parameters from K15 is used to represent the distance distribution of satellite member stars:

\begin{equation}
P_{sat}(r)=\frac{r}{h^2}\,\exp{\left(-\frac{r}{h}\right)}
\end{equation}

\noindent{}where $h$ is the exponential scale length.

The model for the background sources assumes a uniform distribution within the field  $P_{bg}(r)= {2\,r}/{r_f^2}$, where the $r_f$ is the field radius. 

The full list of parameters in our membership modelling for both Reticulum~2 and Horologium~1 was $f,V_{sat},\sigma_{sat},{\rm [Fe/H]}_{sat},\sigma_{{\rm [Fe/H]},{sat}}$ and $V_{bg},\sigma_{bg},{\rm [Fe/H]}_{bg},\sigma_{{\rm [Fe/H]},{bg}}$, respectively. 

\begin{figure*}
\includegraphics{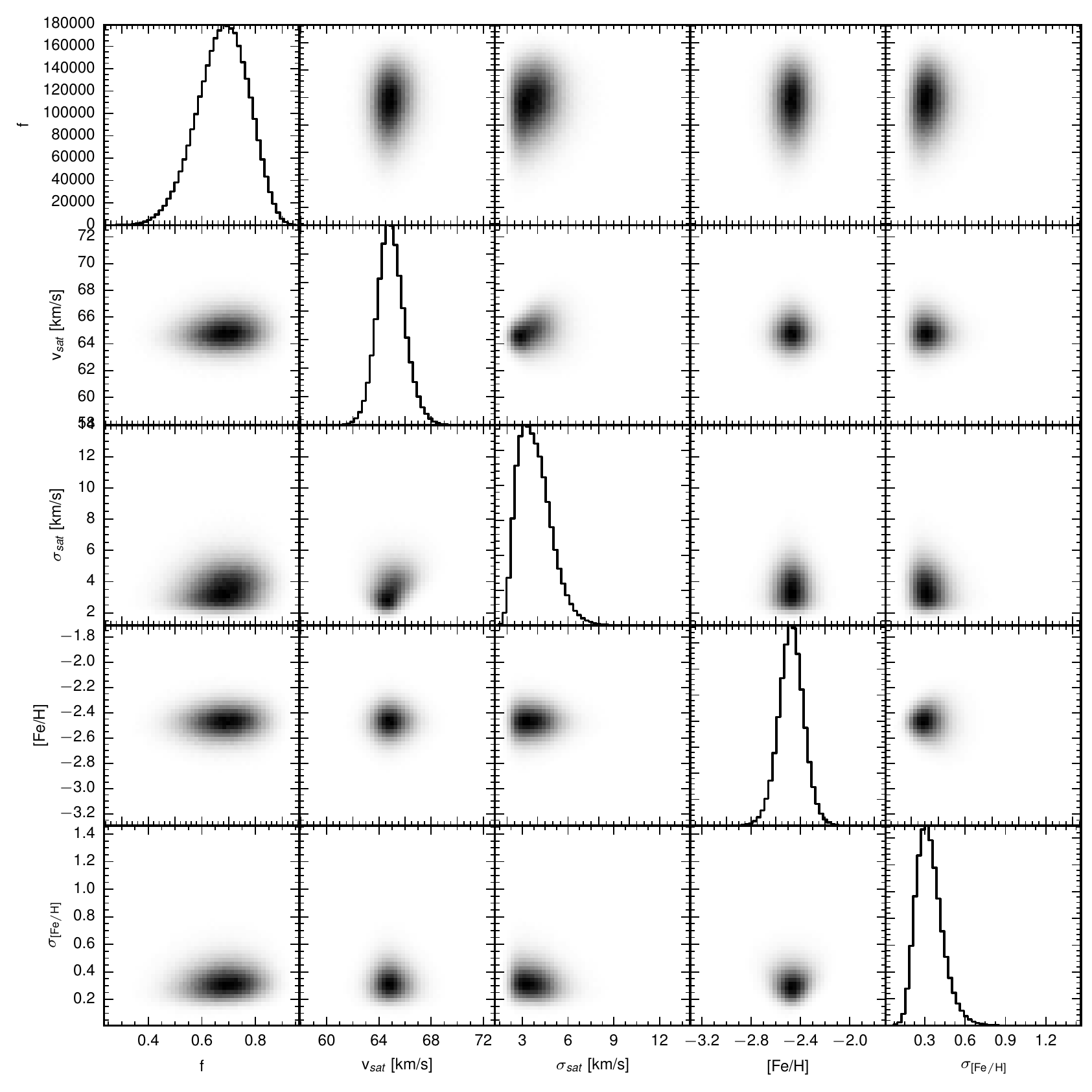}
\caption{2D and 1D marginalized posteriors for the parameters of the chemo-dynamical modeling of Reticulum~2} 
\label{fig:chains_reti2}
\end{figure*}

\begin{figure*}
\includegraphics{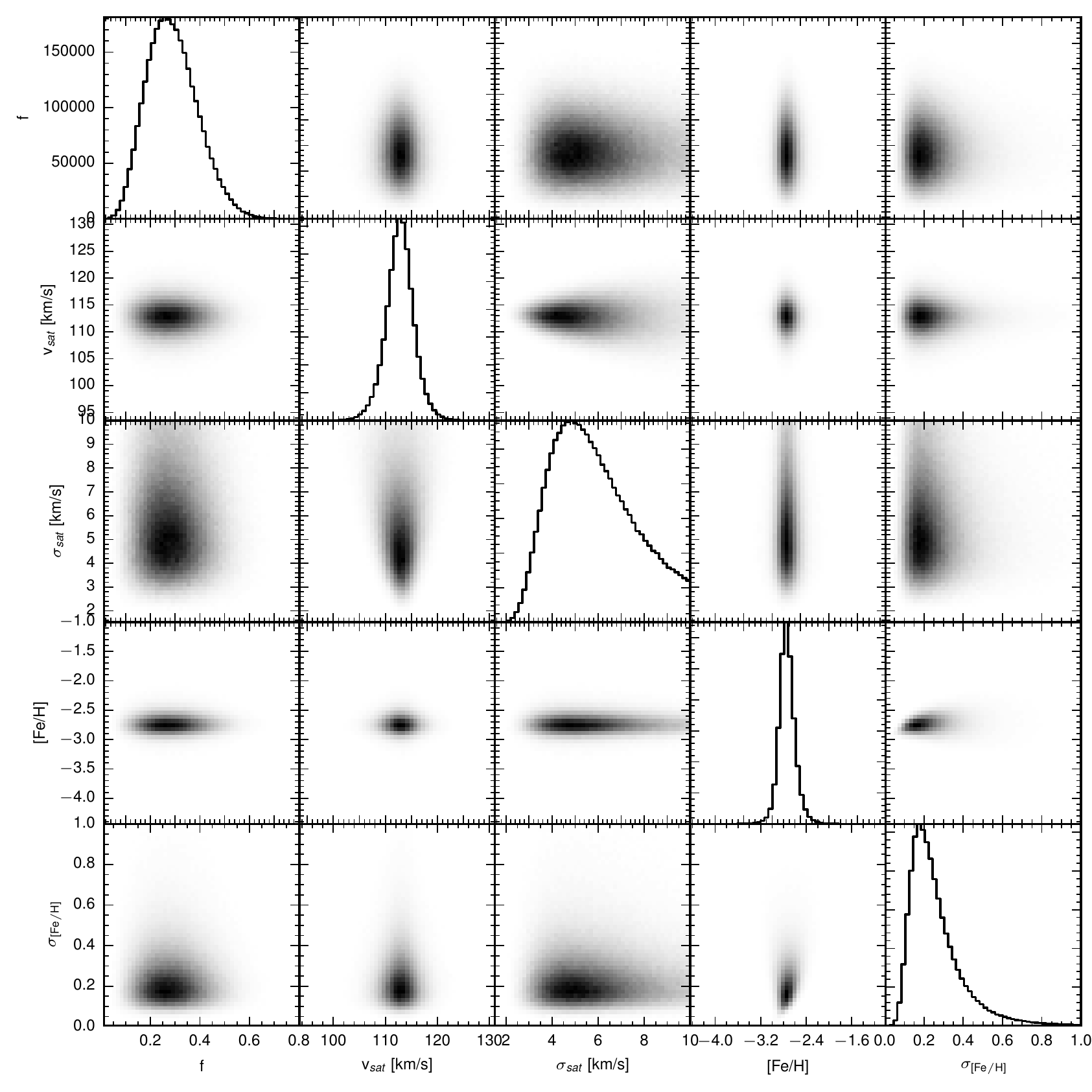}
\caption{2D and 1D marginalized posteriors for the parameters of the chemo-dynamical modeling of Horologium~1} 
\label{fig:chains_horo1}
\end{figure*}

We adopt uninformative priors on $V_{sat}$, $V_{bg}$, ${\rm [Fe/H]}_{sat}$ and ${\rm [Fe/H]}_{bg}$, Jeffreys priors on the distribution dispersions and $f$. The posterior was then sampled using the ensemble MCMC sampler implemented in \texttt{Python} by \citet{foreman_mackey_2013}. The posteriors for the parameters of the satellites are shown on Figures~\ref{fig:chains_reti2} and \ref{fig:chains_horo1}. The resulting parameter measurements quoted in Table~\ref{tbl-1} are the 1D MAP values, and the uncertainties are the 68\% percentiles.


\begin{figure*}
\includegraphics{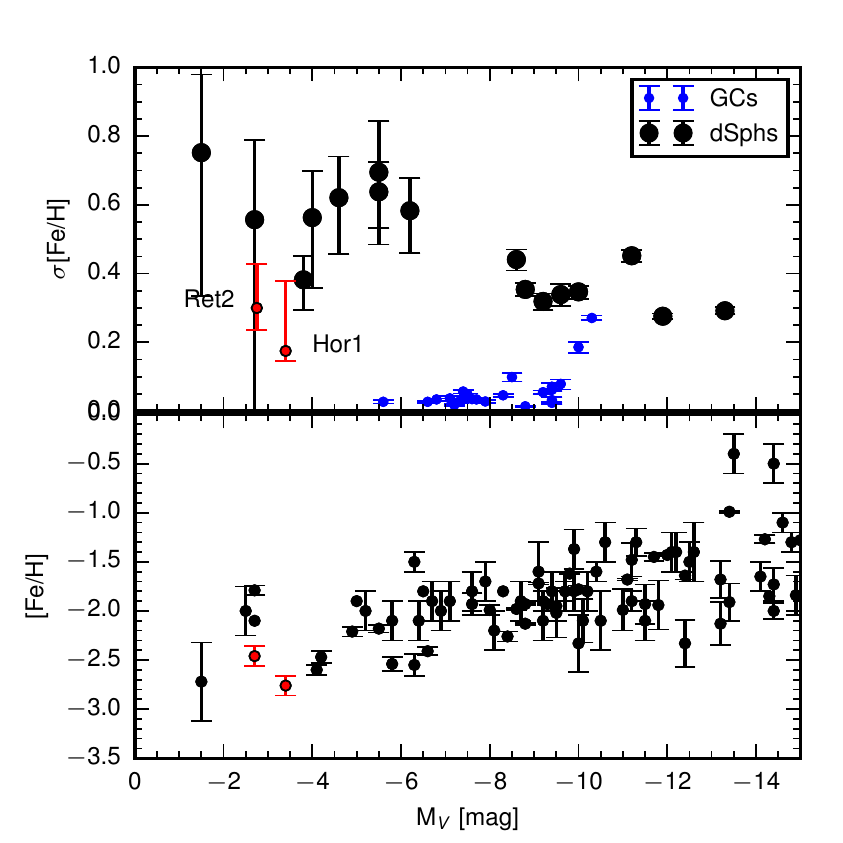}
\includegraphics{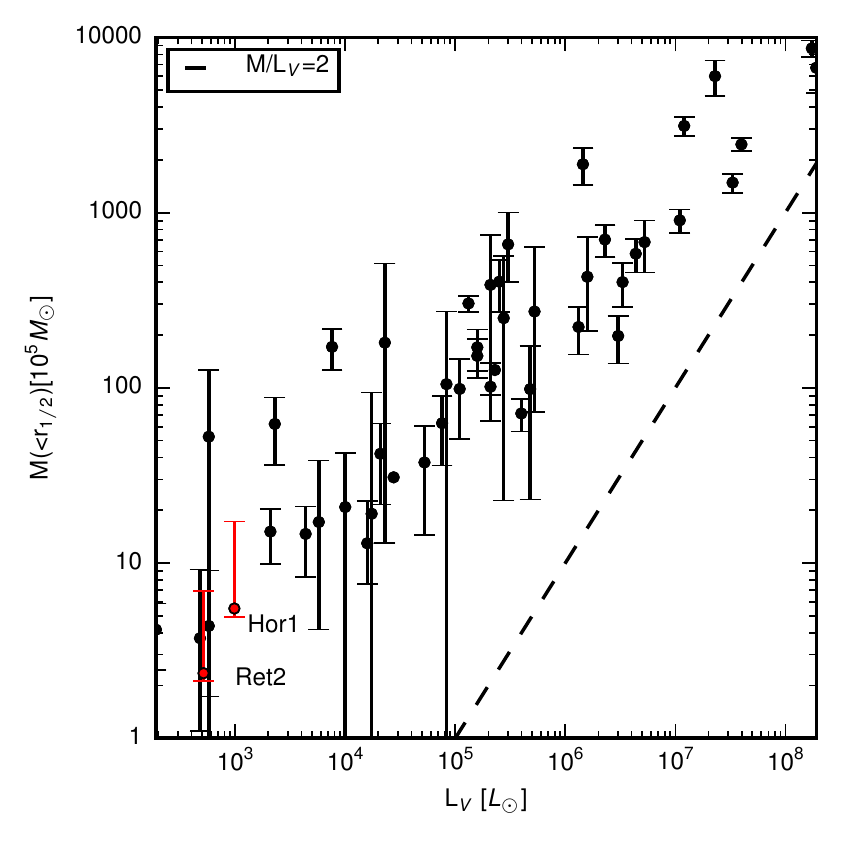}

\caption{{\it Left panel.} Mean metallicity (bottom panel) and metallicity spread (top panel) as a function of luminosity for faint stellar systems. The data from \citet{mccon12,willman_2012} were used. The measurements of Horologium~1 and Reticulum~2 are shown by red circles. Our measurements confirm that these recently discovered galaxies extend the existing mass metallicity trends, and show significant spreads in abundances similar to the ones observed among other dwarf galaxies and not in globular clusters. {\it Right panel} The dynamical mass inside half-light radius vs galaxy luminosity. Both Reticulum~2 and Horologium~1 are on the very faint end of the existing relationship. The dashed line shows the dynamical mass expected for the stellar population with M/L$_V$=2 (i.e., without dark matter).}
\label{fig:feh_scatter}
\end{figure*}

\section{Discussion and Conclusions}
\label{sec:discussion}

\subsection{Reticulum 2}
On the basis of inferred kinematics and chemistry, our analysis has unambiguously identified 18 members in Reticulum~2. Of these stars, 17 are red giants and one is a horizontal branch star. 

We find an intrinsic velocity dispersion  of $3.22^{+1.64}_{-0.49}$~km\,s$^{-1}$ in
Reticulum~2. Although our value is slightly lower than the parallel analyses by \citet{walker_2015b} and \citet{simon_2015}\footnote{\citet{simon_2015} use the Gaia-ESO data as well in their analysis of Reticulum~2}, the velocity
dispersion measurements from all studies are consistent within the uncertainties. As already pointed out by \citet{simon_2015} and \citet{walker_2015b}, the velocity dispersion unambiguously indicate that Reticulum~2 is a dwarf galaxy. Using the mass estimator of \citet{wolf10} we can estimate the total mass inside half-light radii to be  $2.54^{+4.52}_{-0.32}\times 10^5 M_\odot$ for Reticulum~2, which corresponds to a mass-to-light ratio of $\sim$ 500. However the total mass and the mass-to-light ratio has to be treated with caution, as Reticulum~2 is very elongated (axis ratio of 0.4) and is potentially being tidally disrupted (see K15), therefore the mass estimator could be significantly biased.

We also find a substantial spread in overall metallicity of 
$\sigma\left({\rm [Fe/H]}\right) = 0.29^{+0.13}_{-0.05}$\,dex. In contrast to 
\citet{simon_2015}, we have also inferred $[\alpha/{\rm Fe}]$ abundance ratios 
for all satellite candidates observed through the Gaia-ESO Survey. Although the 
uncertainties on $[\alpha/{\rm Fe}]$ are large for most of our confirmed members,
we found the Reticulum~2 data tended towards high $[\alpha/{\rm Fe}]$ ratios. 
Indeed, the lowest $[\alpha/{\rm Fe}]$ ratio of our 18 confirmed members exceeds
$+0.2$\,dex. The resulting estimate of $[\alpha/{\rm Fe}]$ for Reticulum~2 is 
$[\alpha/{\rm Fe}] = 0.40 \pm 0.04$, consistent with observations of 
well-studied present-day dwarf galaxies \citep{tolstoy_2009,kirby_2011}.

There are slight discrepancies in the estimated mean metallicity of Reticulum~2
between this study and \citet{simon_2015}. \citet{walker_2015b}
finds ${{\rm [Fe/H]} = -2.67^{+0.34}_{-0.34}}$, consistent with our measurement
of ${{\rm [Fe/H]} = -2.46^{+0.09}_{-0.10}}$. 
\citet{simon_2015} finds a comparable value of ${\rm [Fe/H]} = -2.65 \pm 0.07$
from \ion{Ca}{2} equivalent widths.
The quoted uncertainty by \citet{simon_2015} are the lowest of all studies, but given
our uncertainties, ${\rm [Fe/H]} = -2.65$ is a mere $1.9\sigma$ deviation.
We explored this possible discrepancy by searching the Gaia-ESO Survey for 
HR10 and HR21 spectra of HD~122563, a 
well-studied metal-poor giant star. HD~122563 is a Gaia benchmark
star \citep{jofre_2014}, and has atmospheric parameters comparable to the stars
in Reticulum~2. The lowest S/N in any single exposure of HD~122563 was $\sim$20. We analysed these data with the model described
in Section \ref{sec:chemical-analysis}, except we found it necessary to use a fourth
order polynomial to account for the continuum in the HD~122563 spectra. 
We find a MAP ${\rm [Fe/H]} = -2.79$, in good agreement with the accepted literature value of ${\rm [Fe/H]} = -2.64$ \citep{jofre_2014}. If anything our metallicity scale may be $\sim$0.1\,dex
more metal-poor than the benchmark, the
opposite direction to the discrepancy in \citet{simon_2015}. Nevertheless,
our robust uncertainties make our measurement reasonably consistent with 
\citet{walker_2015b} and \citet{simon_2015}.

An unexpected discovery was also made in the Gaia-ESO Survey Reticulum~2 data.
The left panels of Figure \ref{fig:stellar_params} show the inferred radial
velocity and metallicity from the model described in Section \ref{sec:chemical-analysis}.
Two stars (Reti~0 and Reti~13) are present at ${V_{hel} \sim 219}$~km\,s$^{-1}$ 
with indistinguishable metallicities of ${{\rm [Fe/H]} \sim -2}$. 
The $\log{g}$ is largely uninformative
for these stars: the MAP values are consistent with a sub-giant star, but the
uncertainties are sufficiently large that both a dwarf or sub-giant are equally
plausible. We checked the \citet{walker_2015b} and \citet{simon_2015} studies
for other stars at comparable velocities. We found one match in \citet{walker_2015b} (star Ret2-153 in their nomenclature), which turned out to be Reti~0, and unsurprisingly both
stars were in the \citet{simon_2015} study, simply marked as `non-members' of 
Reticulum~2. A subsequent search in the surrounding Gaia-ESO Survey Milky
Way field (e.g., non-Reticulum~2 candidates that were in the same field)
revealed a further two stars $(\alpha,\delta)=(53.69300,-54.17860)$ and $(53.73709,-54.10720)$ with similar systemic velocities to Reti~0
and Reti~13 (223.7~km\,s$^{-1}$, 221.3~km\,s$^{-1}$). We have not inferred stellar parameters for these additional two stars. 
However, if we ignore the information that Reti~0 and Reti~13 have
indistinguishable metallicities, a simple calculation using the average number of stars per km\,s$^{-1}$ at RV $\sim{}$200~km\,s$^{-1}$ gives the significance of having 4 stars within $\sim$5~km\,s$^{-1}$ to be $\sim$99.5\% \citep[after correcting for ``look-elsewhere'' effect, e.g. ][]{gross10}. Analogous to the 300\,km s$^{-1}$ stream near Segue~1 
\citep{geha_2009, norris_2010, frebel_2013}, this kinematic feature is 
yet another reminder of the highly substructured nature of the Milky Way
halo \citep[e.g.,][]{schlaufman_2009, starkenburg_2009}.

\subsection{Horologium 1}
With only 5 confirmed members in Horologium~1, we are far more sensitive
to stochastic sampling effects than what we are for Reticulum~2. Nevertheless,
we find a large kinematic dispersion of  $4.9^{+2.8}_{-0.9}$\,km s$^{-1}$
and a metallicity dispersion of $0.17^{+0.2}_{-0.03}$, firmly grouping
Horologium~1 with other known dwarf galaxies. Note that our posteriors on
$\sigma\left(V\right)$ and $\sigma\left({\rm [Fe/H]}\right)$ have 
considerable asymmetry towards higher velocity and metallicity dispersions,
which is primarily attributable to the low number of confirmed members in
our sample. When more data becomes available, it is reasonable to expect 
that a larger metallicity dispersion may be found for Horologium~1, as our MAP
$\sigma\left({\rm [Fe/H]}\right)$ is the lowest reported measurement for
comparable ultra-faint dwarf galaxies (Figure \ref{fig:feh_scatter}). 
A lower metallicity dispersion is strongly disfavoured by our data, and would
be inconsistent with the large velocity dispersion we observe. We also find
Horologium~1 to have $[\alpha/{\rm Fe}] = 0.30 \pm 0.07$, consistent with
the Milky Way dwarf galaxy population.

According to the mass estimator of \citet{wolf10} we estimate the total mass inside the half-light radii of Horologium~1 is $5.25^{+11.5}_{-0.78} \times 10^5 M_\odot$ (notice however a very big error bar). The mass-to-light ratio of $\sim$600 is similar to that observed in Reticulum~2.

\subsection{Comparison to other ultra-faints}

Overall both Reticulum~2 and Horologium~1 systems seem to be quite representative samples of the other known ultra-faint dwarf galaxies. 
\begin{itemize} 
\item The average metallicity of stars in both systems is very low (one of the lowest among dwarf galaxies), but both dwarfs lie well on the existing mass/luminosity - metallicity correlation (see bottom left panel  of Figure~\ref{fig:feh_scatter}). 

\item The metallicity spread, although uncertain, is significantly different from zero, matching what is observed in other dwarf galaxies (top left panel of Figure~\ref{fig:feh_scatter}). It is possible though that the spread seen in Horologium~1 and Reticulum~2 is somewhat smaller than the spread of 0.5--0.7\,dex observed in other ultra-faint systems such as Segue~1 \citep{simon_2011}, but this could simply be a result of small sample sizes.

\item The dark matter content and the mass-to light ratio in the observed systems seem to agree well with the existing correlations with galaxy luminosity (right panel of Figure~\ref{fig:feh_scatter}). 
This suggests that even with these new discoveries of ultra-faint dwarfs we have not reached the limiting density scale of dark matter, which would inform us about the elusive properties of dark matter \citep{gilmore_2007}.
\end{itemize}

\subsection{Possible association with the Magellanic Clouds}

Given their proximity to the LMC and the SMC on the sky, there exists
an exhilarating possibility that some of the newly discovered
satellites, including Reticulum~2 and Horologium~1, have once been part of the
Great Magellanic Family. If such a connection proves true, there is
hope to link the internal properties of the dwarfs (e.g., their dark
matter content, the star-formation and the enrichment histories) with
their orbital motion before and during the accretion onto the Milky
Way. Thus, finally, an in-depth self-consistent picture of the UFS
formation and evolution can be assembled.

Considering the total number of satellites discovered in the SDSS, VST ATLAS and PanSTARRs surveys, the relatively small patch of sky covered by the first year of DES observations appears unusually rich in satellites.
According to
K15, the over-density of satellites around the
Magellanic Clouds is moderate but significant, with at least 3-4
objects possibly belonging to the LMC/SMC pair. Note however, that
the above calculation does not account for the fact that for the faintest systems (e.g., direct analogs of Reticulum~2 and Horologium~1), the SDSS census is incomplete beyond 50 kpc. Therefore the number of
faintest dwarfs within the DES footprint -- namely those with $M_V>-4$ -- 
have to be estimated under the assumption of their Galacto-centric
radial distribution. Given perfect freedom, it seems plausible to
find a radial profile flat enough to produce as many faint
satellites as have been discovered in the DES data. This, however,
would seem to be in tension with the lack of discoveries from surveys
like VST ATLAS and PanSTARRs. In the absence of completeness
estimates for the ongoing imaging surveys, we attempt to clarify the
connection between the newly discovered satellites and the Magellanic
Clouds by complementing their 3D positions with the radial velocity
measurements obtained with the VLT.

The satellites' kinematics can be compared to predictions made from 
osmological zoom-in simulations. For example, according to
\citet{sales_2011}, the distribution of the satellites in
phase-space reveals the time of accretion of the Magellanic system. By
finding one suitable LMC analog in the high-resolution Aquarius suite
\citep{springel_2008}, \citet{sales_2011} convincingly demonstrate
that a high concentration of the former LMC companions is expected in
the Cloud's vicinity if the LMC has only had one peri-center
crossing. More recently, a systematic analysis of 25 LMC analogs in the
ELVIS suite of cosmological zoom-in simulations has been presented by
\citet{deason_2015}. Here, rather than a report on a case study, an
evidence for a trend between the $z=0$ phase-space scatter of the LMC
satellites' and the group in-fall time is presented. According to
\citet{deason_2015}, the observed distribution of satellites
discovered in the DES data appears consistent with a recent, i.e. $<$2
Gyr, accretion. For such late events, the authors also provide a rough
estimate of the total number of past LMC satellites in the DES
footprint: based on positions alone, there should be at least 4 such
objects in the current DES sample. Both \citet{sales_2011} and
\citet{deason_2015} emphasize the role that kinematics play in
uncovering the origin of the Milky Way satellites: chance spatial
alignments are possible, however these are in general less likely in
the vicinity of the group's central (e.g., in the LMC).

The benefit of these N-body simulations is that they paint a fully consistent cosmological portrait of the Magellanic Group, both in terms of accretion history as well as the amount of expected substructure.
However, it is obvious that these simulations
cannot match either the exact orbit of the LMC nor the
presence of its massive companion, the SMC. To complement the
cosmological N-body zoom-in runs, controlled simulations of LMC/SMC
accretion can be mass-produced for a much larger range of the in-fall
parameters \citep[e.g.,][]{nichols_2011}. We will describe the outcome
of such an experiment in the future (see Jethwa et al., in
prep.). Meanwhile, we can shed some light onto possible links between the Magellanic Clouds and Reticulum~2 or Horologium~1 by comparisons with the observed kinematics of the Magellanic gaseous Stream (MS).
The stream
of neutral hydrogen emanating from the Clouds has been mapped out
across tens of degrees, and is complemented with well-documented kinematics 
\citep[see e.g.,][]{putman_2003, nidever_2008}. Additionally, several
numerical models exist explaining the genesis of the MS \citep[see
  e.g.,][]{besla_2010, diaz_2012}.
  
  \begin{figure*}
\begin{center}
\includegraphics[width=0.94\textwidth]{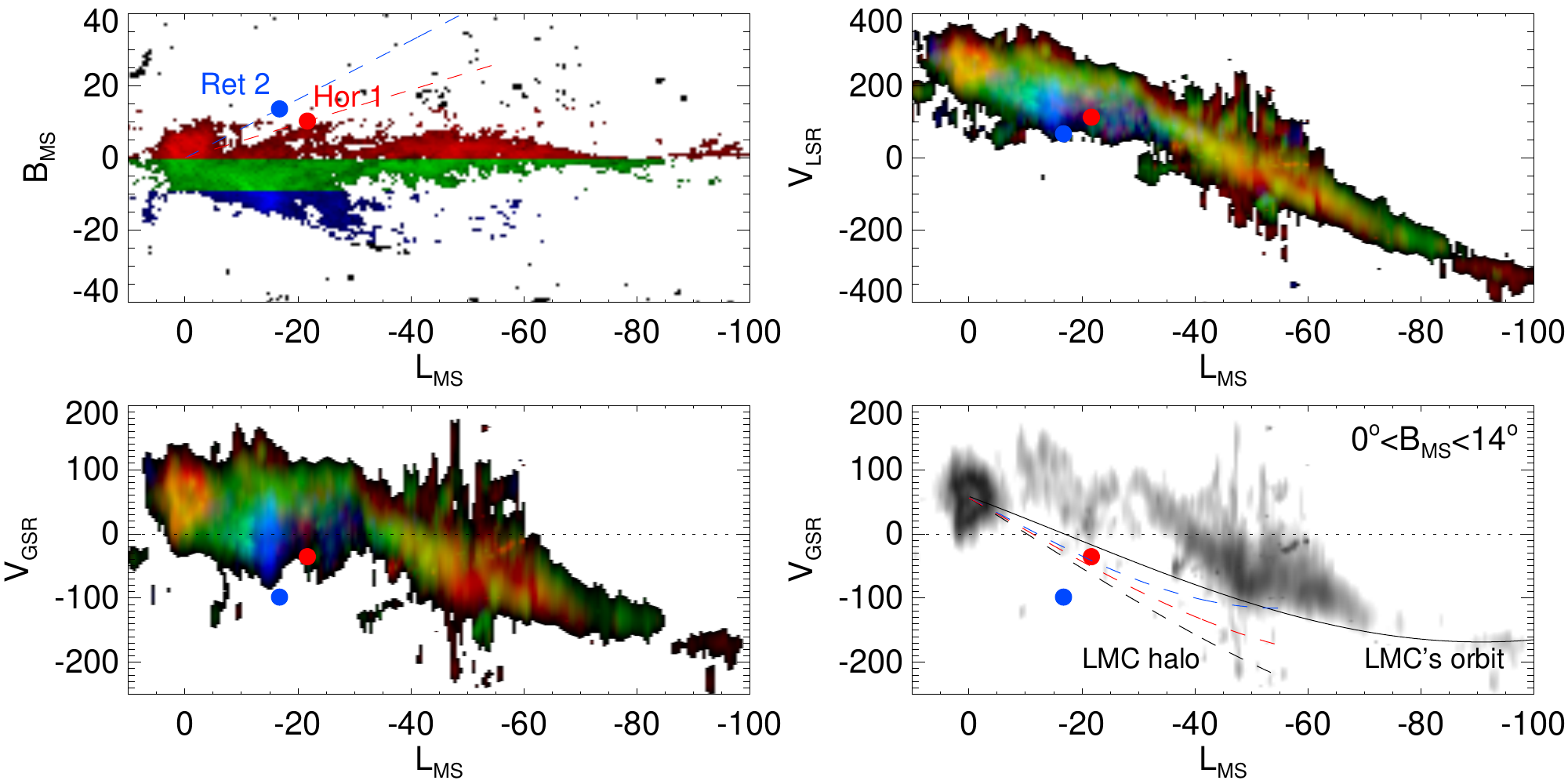}
\end{center}
\caption{Comparison of the kinematics of Reticulum~2, Horologium~1 and the
  Magellanic Stream. {\it Top Left:} Column density of \ion{H}{1} in the MS as detected by \citet{nidever_2008} projected in the
  MS coordinate system ($L_{\rm MS}$, $B_{\rm MS}$). For further
  exploration, the gas is split into three bins in latitude $B_{\rm
    MS}$, shown in red ($0^{\circ}<B_{\rm MS}<14^{\circ}$), green
  ($-9^{\circ}<B_{\rm MS}<0^{\circ}$) and blue ($-23^{\circ}<B_{\rm
    MS}<-9^{\circ}$) color. The location of Reticulum~2 (Horologium~1) are shown by the
  blue (red) filled circle. Blue (red) dashed line marks the locations of
  the LMC halo chosen for the line-of-sight velocity predictions
  that are displayed in the bottom right panel. {\it Top Right:} Heliocentric
  radial velocity $V_{\rm LSR}$ as a function of the MS longitude
  $L_{\rm MS}$. The MS \ion{H}{1} gas density in this phase-space projection is
  shown as a false-color RGB composite built with grey-scale density
  distributions from corresponding $B_{\rm MS}$ bins shown in the left
  panel. Note the strong $V_{\rm LSR}$ velocity gradient, which is
  chiefly caused by the Solar reflex motion. Both Reticulum~2 and Horologium~1
  appear to lie close to the lower envelope of the MS velocity
  signal. {\it Bottom Left:} Same as the top right panel, but for
  Galactocentric radial velocity $V_{\rm GSR}$ as a function of the MS
  longitude $L_{\rm MS}$. Note that the transformation from the LSR to
  the GSR has significantly reduced the velocity gradient observed in
  the top right panel. While the velocities of Reticulum~2 and Horologium~1 are
  consistent with the lower range of the MS stream kinematics at the
  corresponding $L_{\rm MS}$, the gas with comparable $V_{\rm GSR}$ is
  located mostly at $B_{\rm MS} < 0^{\circ}$, evidenced by the blue
  color of the region of the MS nearest to the satellites. This is further
  illustrated in the bottom right panel. {\it Bottom Right:} Motions of
  the MS gas with $0^{\circ}<B_{\rm MS} < 14^{\circ}$. This map
  confirms that near Reticulum~2, the MS \ion{H}{1} gas motions are different by
  approximately 100~km\,s$^{-1}$. On the hand, near Horologium~1, there exists
  MS \ion{H}{1} gas whose velocity is similar to that of the dwarf. The black
  solid line is the track of the LMC's orbit (see K15). Dashed curves show the prediction for mean
  velocity of the LMC halo, i.e. the projection of the LMC's velocity
  vector onto the line-of-sight. Black dashed curve corresponds to the
  lines of sight crossing the LMC halo at $L_{\rm MS}=0^{\circ}$, and the blue
  and red dashed curves correspond to the halo slices shown in the top
  and left panels. Reticulum~2 is $\sim 80$~km\,s$^{-1}$ away from the halo
  prediction. However, Horologium~1 is a mere $\sim$15~km\,s$^{-1}$ away from the halo prediction.}
\label{fig:mstream}
\end{figure*}

Figure~\ref{fig:mstream} shows the positions of the two satellites in
the space of MS longitude $L_{\rm MS}$, MS latitude $B_{\rm MS}$ and
radial velocity. The top left panel of the figure gives the locations of Reticulum~2
and Horologium~1 with respect to the distribution of the column density of \ion{H}{1}
gas in the Magellanic stream as detected by \citet{nidever_2008}. We
split the \ion{H}{1} detections into three bins according to the latitude
$B_{\rm MS}$, marked with red ($0^{\circ}<B_{\rm MS}<14^{\circ}$),
green ($-9^{\circ}<B_{\rm MS}<0^{\circ}$) and blue
($-23^{\circ}<B_{\rm MS}<-9^{\circ}$) colors. The location of Reticulum~2
(Horologium~1) is also shown as blue (red) filled circle. The top right panel
presents the false-RGB composite map of \ion{H}{1} in the plane of the MS
longitude $L_{\rm MS}$ and heliocentric radial velocity $V_{\rm
  LSR}$. As previously shown by \citet{nidever_2008}, in this projection the Stream's \ion{H}{1} content forms a broad band, typically $\sim$100~km\,s$^{-1}$ in
extent. As indicated by the rapidly changing color, portions of the
stream at varying $B_{\rm MS}$ contribute different amounts of
velocity signal at given $L_{\rm MS}$. Note however that the steep
velocity gradient as a function of the MS longitude is predominantly
caused by the Solar reflex motion. This is confirmed in the bottom
left panel of Figure~\ref{fig:mstream} which displays the map of \ion{H}{1} in the plane of
$L_{\rm MS}$ and Galacto-centric radial velocity $V_{\rm GSR}$. The
$V_{\rm GSR}$ signature of the stream remains largely constant up to
$L_{\rm MS}\sim-40^{\circ}$ where it starts to decline, but slower than with respect to $V_{\rm LSR}$.

As obvious from both velocity maps described above, the MS \ion{H}{1} is
detected near the positions of Reticulum~2 and Horologium~1 in the ($L_{\rm MS}$,
$V$) plane. However, the region of phase-space occupied by the
dwarfs is dominated by the gas at negative $B_{\rm MS}$ (i.e., around
the SMC) as evidenced by the blue tint of this portion of the
image. The bottom right panel of Figure~\ref{fig:mstream} illustrates the range of
possible velocities of the MS with positive $B$ only (i.e., those around Reticulum~2 and Horologium~1). At the longitude of Reticulum~2, there is
a gap of order of 100~km\,s$^{-1}$ between the grey-scale density map of
the MS \ion{H}{1} and the velocity of Reticulum~2. However near Horologium~1, the MS
gas completely extends to the measured velocity of the satellite.

The offset in the line-of-sight velocity between the trailing MS \ion{H}{1}
and the satellites is expected, as stripped gas can experience a
drag force through interactions with the hot Galactic corona. The drag
will decelerate the stream clouds, causing them to fall to lower
Galactocentric radii. To make sense of the radial velocities of Reticulum~2
and Horologium~1, let us recall their locations with respect to the Milky
Way center and the Magellanic Clouds. In 3D space, Reticulum~2 is in front of the
Clouds, $\sim$24\,kpc away from the LMC and $\sim$39\,kpc from the SMC,
while Horologium~1 is behind the Clouds, $\sim$38\,kpc away from the LMC and
$\sim$32\,kpc from the SMC. Importantly, both satellites are trailing
the LMC, as evidenced from the Cloud's orbital motion shown in Figure~20 of K15. 

The black solid line in the bottom right panel of Figure~\ref{fig:mstream}
shows the projection of a backward-integrated LMC orbit from
K15, namely the one with the NFW's concentration
$c=10$, projected onto the plane of $L_{\rm MS}$ and $V_{\rm GSR}$. The orbit
attains negative line-of-sight velocities at $L_{\rm MS}<-20^{\circ}$
and appears to be in reasonable agreement with the velocity of Horologium~1
at the corresponding MS longitude. This implies that the 4D coordinate
of Horologium~1 is consistent with those expected for the LMC's trailing
debris. At the location of Reticulum~2 the velocity gap of
$\sim$100~km\,s$^{-1}$ persists. It is tempting to assert that this precludes the possibility of an association between Reticulum~2 and the LMC. However, it is unrealistic to expect Reticulum~2 to behave simply like the LMC trailing debris. During tidal disruption it is normal to expect the trailing debris to form from particles with higher energy and angular momentum than the progenitor. Subsequently the trailing debris also have larger Galactocentric radii on
average, and longer orbital periods. Yet, we know that Reticulum~2 is closer
to the Milky Way's center than the LMC itself. Thus, in order to explain its
origin as part of the Magellanic family, an additional factor needs to
be included. It is conceivable that an interaction with the SMC would be sufficient to drive Reticulum~2 onto it's current orbit.

Finally, we have not yet considered the possibility that Reticulum~2 and/or
Horologium~1 could still be bound to the LMC. Superficially, such situation
seems unlikely given the distances between the LMC and the two
satellites. However, \citet{munoz_2006a} report spectroscopically
confirmed detection of the likely LMC's stellar halo some 22$^{\circ}$
degrees away from the LMC, at angular distances comparable to Reticulum~2 and Horologium~1. Motivated by this discovery, we test whether
the line-of-sight velocities of Reticulum~2 and Horologium~1 are consistent with the LMC's halo. The dashed lines in the bottom right panel of
Figure~\ref{fig:mstream} show projections of the LMC velocity vector
onto several lines-of-sight. The black dashed curve corresponds to the
lines-of-sight slicing the LMC's halo at $L_{\rm MS}=0^{\circ}$. Blue
(red) dashed curve shows the run of the projection of the mean LMC's
halo velocity along the line-of-sight moving from the LMC's center to
the location of Reticulum~2 (Horologium~1) as shown in the top left panel of the
figure. At the position of Reticulum~2, the line-of-sight velocity of
non-rotating LMC's halo would be ${\sim}-20$~km\,s$^{-1}$, some $\sim$80~km\,s$^{-1}$ away the Galactocentric velocity of Reticulum~2, $V^{\rm Ret
  2}_{\rm GSR}\sim -100$~km\,s$^{-1}$. However, at the position of Horologium~1, the LMC's halo's line-of-sight velocity is predicted to be $\sim
-50$~km\,s$^{-1}$, only 15~km\,s$^{-1}$ away from the Galactocentric
velocity of the satellite $V^{\rm Hor 1}_{\rm GSR}\sim -35$~km\,s$^{-1}$. It is important to note that while Horologium~1 is
located at a similar angular distance from the LMC as compared to the
stellar halo detections reported by \citet{munoz_2006a}, it is probably
twice as far away from the LMC in 3D, i.e. 40\,kpc instead of 20\,kpc. To explain the dynamics of the LMC+SMC system and the formation
of the MS, \citet{besla_2010} advocate the existence of a massive
($M_{\rm vir}=3\times 10^{11} M_{\odot}$) dark matter halo of the LMC. The
corresponding virial radius of this halo would be $>$ 100\,kpc, while
its (approximate) tidal radius just under 40\,kpc. If the LMC is
on its first pericenter crossing, then given the weak tides the LMC
has been experiencing so far, it is reasonable to expect that its dark matter
distribution can extend as far as 40\,kpc. Therefore, surprisingly,
there appears to be some probability that Horologium~1 is still
gravitationally bound to the Large Magellanic Cloud.

\acknowledgements

    Based on data products from observations made with ESO Telescopes at
    the La Silla Paranal Observatory under programme ID 188.B-3002. These
    data products have been processed by the Cambridge Astronomy Survey 
    Unit (CASU) at the Institute of Astronomy, University of Cambridge, 
    and by the FLAMES/UVES reduction team at INAF/Osservatorio Astrofisico
    di Arcetri. These data have been obtained from the Gaia-ESO Survey 
    Data Archive, prepared and hosted by the Wide Field Astronomy Unit, 
    Institute for Astronomy, University of Edinburgh, which is funded by
    the UK Science and Technology Facilities Council.
    
    This work was partly supported by the European Union FP7 programme 
    through ERC grant numbers 320360 and 308024 and by the Leverhulme Trust through 
    grant RPG-2012-541. We acknowledge the support from INAF and Ministero 
    dell' Istruzione, dell' Universit\`a' e della Ricerca (MIUR) in the 
    form of the grant "Premiale VLT 2012". The results presented here 
    benefit from discussions held during the Gaia-ESO workshops and 
    conferences supported by the ESF (European Science Foundation) through
    the GREAT Research Network Programme.

    We thank Matt Walker, Mario Mateo, Gurtina Besla, David Nidever, Bertrand Plez, Matthew Ruffoni and Juliet Pickering.
    
    This research made use of Astropy, a community-developed core Python 
    package for Astronomy \citep{astropy_2013}.

\end{document}